\documentclass[fleqn]{2017SCGE}
\def\aa{Astron. Astrophys.}

\def\apj{Astrophys. J.}

\def\apjsup{Astrophys. J. Suppl. Ser.}

\def\jgr{J. Geophys. Res.}
\def\mnras{Mon. Not. R. Astron. Soc.}
\def\pasj{Publ. Astron. Soc. Jap.}

\def\sp{Solar Physics}

\begin{document}
\ensubject{subject}

\ArticleType{Rview}

\title{From Polarimetry to Helicity: Studies of Solar Magnetic Fields at the Huairou Solar Observing Station}{From Polarimetry to Helicity}
\author[]{Hongqi Zhang}{{hzhang@bao.ac.cn}}

\AuthorMark{Hongqi Zhang}

\AuthorCitation{Hongqi Zhang}

\address[]{Key Laboratory of Solar Activity, National Astronomical Observatories of the Chinese Academy of Sciences, 100101, Beijing China}

\abstract{
In this paper, we briefly introduce the basic questions in the measurements of solar magnetic fields and the possible error sources  due to the approximation of the theory of radiation transfer of spectral lines in the solar atmosphere.  We introduce some basic research progress in magnetic field measurement at Huairou Solar Observing Station of National Astronomical Observatories of the Chinese Academy of Sciences, especially concerning the non-potentiality in solar active regions, such as  the magnetic shear, current and helicity.  We also discuss some basic questions for the measurements of the magnetic fields and corresponding challenges for the future studies. 
}



\maketitle
\begin{multicols}{2}  
\section{Introduction}

Since the 80s of the last century, a series of solar vector magnetograms have been observed at  Huairou Solar Observing Station of National Astronomical Observatory of the Chinese Academy of Sciences. These observations present a large mount of  information  on the solar magnetic activities, including the basic structure, evolution of the magnetic field, the formation of the non-potential magnetic field, the current and also the corresponding helicity in solar active regions. 

It is noticed that our research of solar magnetic fields is usually based on 
the theory of radiation transfer in the solar magnetic atmosphere. This implies that the general information of the solar photospheric vector magnetic field and the chromospheric magnetic field is based on the measurement of polarized light by means of the Zeeman effect. The study of the line formation in solar magnetic atmosphere not only provides a basic estimation for understanding the accuracy of the measured solar magnetic field, but also the base of the theoretical analysis of the solar magnetic fields. This leads us need to know the possible sources of errors or deviations obtained from these magnetic field observations, even if the errors caused by optical instruments have been ignored. 

In this paper, we try to review, examine and analyze some background of solar magnetic field research based on   magnetic field data obtained at Huairou Solar Observing Station, and highlighting some scientific achievements. We only briefly discuss some of the topics concerning fundamental questions from observations of solar magnetic fields, and do not give references to all aspects. Nevertheless, there are still some fundamental questions  remained and to be researched in the future.

\section{General Description of Radiative Transfer}

{ The radiative transfer of the magnetically sensitive lines in solar atmosphere is a complex process. The diagnostic of solar magnetic fields based on these magnetically sensitive lines normally accompanies with  some  assumptions and approximations of the theoretical analysis, which probably are some  error sources of the inversion of solar magnetic fields from observations.     
The important is to prevent or avoid the influence of  these errors and trying to obtain relative accurated information of solar magnetic fields. As the relevant forms of the solution of the radiative transfer equations have been used to detect the solar magnetic fields,  the corresponding spectral line formation in solar magnetic fields need to be analyzed in more detail.}


In the stellar magnetic atmosphere, the non local thermodynamic equilibrium (non-LTE) equation of Stokes vector ${\textbf{ I}_\nu}=(I_\nu,Q_\nu,U_\nu,V_\nu)$ for the transfer of polarized radiation can be written in a general matrix form (Stenflo (1994) \cite{Stenflo94}) 
\begin{equation}
\label{eq:transij}
\frac{d\textbf{ I}_\nu}{ds}=-(N_{J_l}B_{J_lJ_u}\Phi_{J_l}-N_{J_u}B_{J_uJ_l}\Phi_{J_u})\frac{h\nu}{4\pi}\textbf{  I}_\nu+\textbf{  j},
\end{equation}
where in the case of non-scattering process the emission vector is approximately by
\begin{equation}
\label{eq:trasj}
\textbf{  j}=\frac{h\nu}{4\pi}\Phi_{J_u}\textbf{  1}N_{J_u}A_{J_uJ_l},    
\end{equation}
and  the relationship with
\begin{eqnarray}
\label{eq:trampar}
&&A_{ji}=\frac{2h\nu^3}{c^2}B_{ji}, ~~~ N_iC_{ij}=N_jC_{ji},\\
&&\frac{C_{ij}}{C_{ji}}=\frac{g_j}{g_i}e^{E_{ij}/kT}~~~\mathrm{and}~~~g_{j}B_{ji}=g_{i}B_{ij},
\end{eqnarray}
where $B_{ij}$, $B_{ji}$ and $A_{ji}$ are Einstein coefficients, $C_{ij}$ and $C_{ji}$ are collisional coefficients \cite{Rutten03,Fang08}.  
Non-LTE population departure coefficients $b_i$ are defined as:
\begin{equation}
\label{eq:souf1}
b_l=N_l/N^{LTE}_l, ~~~ b_u=N_u/N^{LTE}_u
\end{equation}
with $N_{(l,u)}$ the actual population and $N^{LTE}_{(l,u)}$ the Saha-Boltzmann values for the lower and upper
levels, respectively. 
The expanded form of ${  \Phi}$ is {\bf the matrix in the right-hand side of} Eq. (\ref{radtrf}), and
we introduce the source function in the form
\begin{equation}
\label{eq:souf2}
S_\nu=\frac{N_jA_{ji}}{N_iB_{ij}-N_jB_{ji}}.
\end{equation}

When we  study the formation of polarized light in the magnetic field with  Stokes parameters, Unno-Rachkovsky equations of polarized radiative transfer of spectral lines  can be taken the form (Unno (1956) \cite{Unno56} and Rachkovsky (1962a, b) \cite{Rachkovsky62a,Rachkovsky62b})
\begin{equation}
\label{radtrf}
\mu\frac{d}{d\tau_c}\left(\begin{array}{c}
I\\
Q\\
U\\
V
\end{array} \right)=
\left(\begin{array}{cccc}
\eta_{0}+\eta_{I} & \eta_{Q} & \eta_{U} & \eta_{V}\\
\eta_{Q} & \eta_{0}+\eta_{I} &  \rho_{V} & -\rho_{U}\\
\eta_{U} & -\rho_{V} & \eta_{0}+\eta_{I} & \rho_{Q}\\
\eta_{V} & \rho_{U} & -\rho_{Q} & \eta_{0}+\eta_{I}
\end{array} \right)
\left( \begin{array}{c}
I-S\\
Q\\
U\\
V
\end{array} \right).
\end{equation}
The symbols have their usual meanings given by Landi
Degl’Innocenti (1976) \cite{Land76}. In Eqs. ($\ref{radtrf}$), $d\tau_c=-\kappa_cds$, and $\eta_I$, $\eta_Q$, $\eta_U$, $\eta_V$ are Stokes absorption coefficient parameters and $\rho_Q$, $\rho_U$ and $\rho_V$ are related to the magneto-optical effects.  

In the analytical solution of the radiative transfer equations for polarized spectral lines  \cite{Unno56,Rachkovsky62a,Rachkovsky62b}, several limitations are introduced: (a) Uni-dimensional plane-parallel atmosphere with a constant magnetic field; (b) linear dependence of the source function with optical depth; (c) constant ratio of the line and continuous absorption coefficients. The transfer equations for polarized radiation can be solved to give
\begin{eqnarray}  
I&=&S\!_0+\mu S\!_1\triangle^{-1}[(1+\eta_I)((1+\eta_I)^2+\rho_Q^2+\rho_U^2+\rho_V^2)],\nonumber\\
Q&=&-\mu S\!_1\triangle^{-1}[(1+\eta_I)^2\eta_Q+(1+\eta_I)(\eta_V\rho_U-\eta_U\rho_V)\nonumber\\
&&+\rho_Q(\eta_Q\varrho_Q+\eta_U\varrho_U+\eta_V\varrho_V)],\nonumber\\
U&=&-\mu S\!_1\triangle^{-1}[(1+\eta_I)^2\eta_U+(1+\eta_I)(\eta_Q\rho_V-\eta_V\rho_Q)\\
&&+\rho_Q(\eta_Q\varrho_Q+\eta_U\varrho_U+\eta_V\varrho_V)],\nonumber\\
V&=&-\mu S\!_1\triangle^{-1}[(1+\eta_I)^2\eta_V+\rho_V(\eta_Q\varrho_Q+\eta_U\varrho_U+\eta_V\varrho_V)]\nonumber.\label{eq:absop} 
\end{eqnarray}
where
\begin{eqnarray*}
\triangle&=&(1+\eta_I)^2[(1+\eta_I)^2-\eta_Q^2-\eta_U^2-\eta_V^2+\rho_Q^2+\rho_U^2+\rho_V^2]\\
&&-(\eta_Q\varrho_Q+\eta_U\varrho_U+\eta_V\varrho_V)^2,
\end{eqnarray*}
and
where the source function $S(\tau)=S\!_0+S\!_1\tau$.

In the approximation of the weak magnetic field (Stix, 2002) \cite{Stix02}, one can obtain the following formulas   
\begin{eqnarray}
\eta_I&=&\eta_p+O(v_b^2),\nonumber\\ 
\eta_Q&=& - \frac{1}{4}\frac{\partial^2\eta_p}{\partial v^2}v_b^2\sin^2\psi \cos2\varphi +O(v^4_b) ,\\
\eta_U&=&- \frac{1}{4}\frac{\partial^2\eta_p}{\partial v^2}v_b^2\sin^2\psi \sin2\varphi +O(v^4_b) ,\nonumber\\
\eta_V&=&\frac{\partial\eta_p}{\partial v}v_b\cos\psi +O(v^3_b),\nonumber\label{eq:weakapprox}
\end{eqnarray}
where
$v_b=\bigtriangleup\lambda_b/\!\bigtriangleup\!\lambda_D$, 
and $\bigtriangleup\lambda_b$ is the corresponding splitting of different subcomponents of the magneto-sensitive spectral lines due to Zeeman effects and $\bigtriangleup\lambda_D$ is the Doppler width.

With Eq. (\ref{eq:absop})  and the neglect  of  magneto-optic effects,  we can get the simple relationship between the magnetic field and Stokes parameters. It is found that
\begin{eqnarray}
I&\approx&S\!_0+\frac{\mu S\!_1}{(1+\eta_I)},\nonumber\\ 
Q&\approx&-\frac{\mu S\!_1}{(1+\eta_I)^2}\eta_Q\approx C_T'B^2\sin^2\psi \cos2\varphi =C_T'B^2_T \cos2\varphi,\nonumber\\
U&\approx&-\frac{\mu S\!_1}{(1+\eta_I)^2}\eta_U\approx C_T'B^2\sin^2\psi \sin2\varphi  =C_T'B^2_T \sin2\varphi ,\nonumber\\
V&\approx&-\frac{\mu S\!_1}{(1+\eta_I)^2}\eta_V\approx C_L'B\cos\psi =C_L'B_L,  
\label{eq:weakapproxa}
\end{eqnarray}
where 
\begin{eqnarray}
C_T'    &=&  \frac{\mu S\!_1}{4(1+\eta_I)^2}\frac{\partial^2 \eta_p}{\partial v^2} \left[\frac{e\lambda_0^2}{4\pi m_ec^2\!\bigtriangleup\!\lambda_D}(M_1g_1-M_2g_2)\right]^2,\\
\displaystyle C_L'    &=&- \frac{\mu S\!_1}{(1+\eta_I)^2}\frac{\partial \eta_p}{\partial v} \frac{e\lambda_0^2}{4\pi m_ec^2\!\bigtriangleup\!\lambda_D}(M_1g_1-M_2g_2).  \label{eq:weakfv} 
\end{eqnarray}
This means that the longitudinal component of magnetic field can be written in the form
\begin{equation}
B_L=C_L V,\label{eq:weekfild1} 
\end{equation}
where $C_l$ is the calibration parameter of longitudinal component of field and is a function of the wavelength of a spectral line. The intensity $B_T$ and azimuth angle $\varphi$ of the transverse
magnetic field can be obtained
\begin{equation}
B_T=C_T \sqrt[4]{Q^2+U^2},\\~~~~~~~~~~
\varphi=\frac{1}{2}\textrm{tg}^{-1}\left(\frac{U}{Q}\right).\label{eq:weekfild2}
\end{equation}

\section{Measurements of Photospheric Magnetic Field}

The studies of  radiative transfer of spectral lines without consideration of scattering in solar magnetic field have been achieved, such as by Ai, Li \& Zhang (1982), Jin \& Ye (1983), Zhang (1986), Song et al. (1990, 1992) and Qu et al. (1997) \cite{Ai82,Jin83,Zhang86,Song90,Song92,Qu97} in China. 

\begin{figure}[H]
\centerline{\includegraphics[width=65mm,angle=0.0]{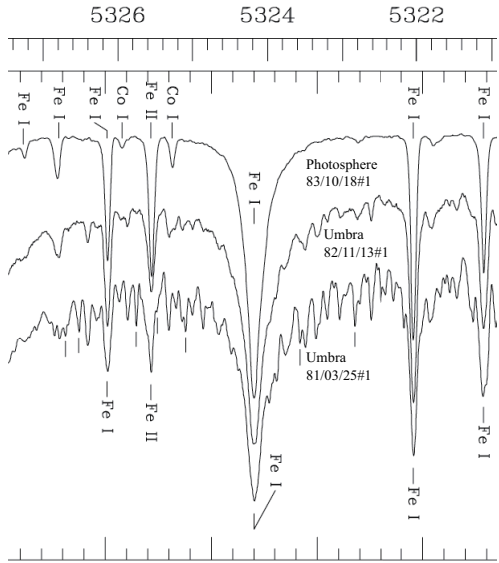}}
\caption{The spectra of FeI 5324.19 {\AA} obtained from Kitt Peak
observational results. From \cite{Kurucz84}   
\label{fig:line5324L} }
\end{figure}

\subsection{Formation of FeI$\lambda$5324.19\AA{} Line in Photospheric Magnetic Field}

The FeI$\lambda$5324.19\AA{} line is a working line  for the Huairou Vector Magnetograph (Solar Magnetic Field Telescope - SMFT) in National Astronomical Observatories of China (Ai \& Hu, 1986) \cite{Ai86}. The FeI$\lambda$5324.19\AA{}  line in Figure  $\ref{fig:line5324L}$ is a normal triplet in the magnetic field and the Lande factor g=1.5, the excitation potential of the low energy level of this line is 3.197eV. The equivalent width of the line is 0.33\AA{} and the residual intensity at the core is 0.17 (Kurucz et al., 1984) \cite{Kurucz84}. This  is a relatively wide line than other normal photospheric lines, such as  the working lines used at the ground-based vector magnetographs, such as the FeI$\lambda$5250.2\AA{} line (Lande factor g=3) used by the Video Vector Magnetograph at Marshall Space Flight Center (West \& Hagyard， 1983) \cite{West83}  and the FeI$\lambda$ 6301.5\AA{} (Lande factor g=1.667) and FeI$\lambda$ 6302.5\AA{} (Lande factor g=2.5) lines used by the Solar Polarimeter at Mees Solar Observatory (Ronan
et al., (1987) \cite{Ronan87}). 

The  bandpass of the birefringent filter of the Huairou Magnetograph with 3 sets KDP crystal modulators is about  0.15\AA{} for the FeI$\lambda$5324.19\AA{} line. The center wavelength of the filter can be shifted  and is normally at -0.075\AA{} from the line center of FeI$\lambda$5324.19\AA{} for the measurement of longitudinal magnetic field and at the line center for the transverse one \cite{Ai86}. The theoretical calibration of Huairou Magnetograph was presented by Ai, Li \& Zhang (1982) \cite{Ai82} and observational one by Wang, Ai, \& Deng (1996) \cite{WangTJ96} and Su \& Zhang (2004) \cite{Su04}. 

\subsection{Numerical Calculation of Radiative Transfer of Stokes Parameters}

The theoretical analysis of Stokes parameters in the magnetic field
is  important for the diagnostic of the magnetic field in the
solar atmosphere. Now, we firstly study the Stokes profiles of the
FeI$\lambda$5324.19\AA{} line  
under the VAL-C quiet Sun model atmosphere (Vernazza et al., 1981) \cite{Vernazza81} by the numerical calculation code of Ai, Li \& Zhang (1982) \cite{Ai82} and fit the observed profile  $I$ (Kurucz, et al., 1984 \cite{Kurucz84}) in the case of no magnetic field. 

\begin{figure}[H]
\centerline
{\includegraphics[width=65mm]{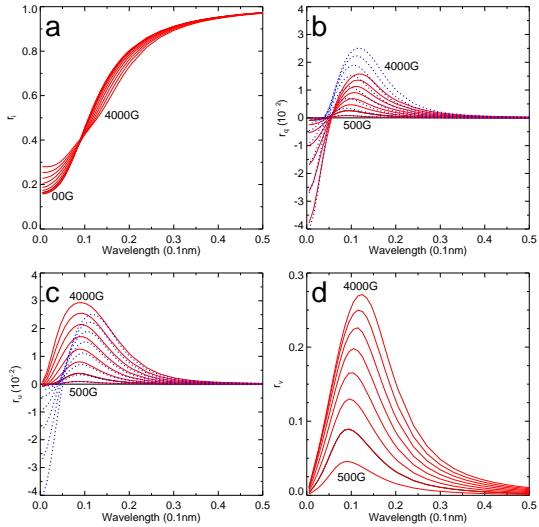}}
\vspace{0.2cm}
\caption{The Stokes parameters $r_i,r_q,r_u$, and $r_v$ (i.e. $I/I_c$, $Q/I_c$, $U/I_c$  and $V/I_c$) of the FeI$\lambda$5324.19\AA{} line  numerically calculated under the VAL-C model atmosphere (red solid lines) \cite{Vernazza81}  with  the inclination $\psi$=30$^\circ$, azimuth $\varphi=22.5^\circ$, and $\mu =1$ for the strength from 500 to 4000 gauss with the interval of 500 gauss, but {\bf Stokes $I/I_c$ from 0 Gs}. The blue dotted lines (in Figures b and c) mark $r_q$ and $r_u$ in the case that the magneto-optical effect has been ignored, respectively.
\label{fig:5324stokes}}
\end{figure}

\begin{figure}[H]
\centerline
{\includegraphics[width=65mm]{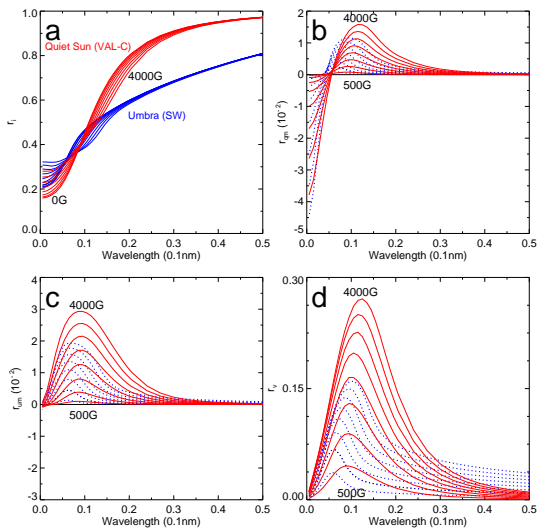}}
\vspace{0.2cm}
\caption{The Stokes parameters $r_i,r_q,r_u$, and $r_v$ (i.e. $I/I_c$, $Q/I_c$, $U/I_c$  and $V/I_c$) of the FeI$\lambda$5324.19\AA{} line  numerically calculated under the VAL-C model atmosphere (red solid lines) \cite{Vernazza81}  with  the inclination $\psi$=30$^\circ$, azimuth $\varphi=22.5^\circ$, and $\mu =1$ for the strength from 500 to 4000 gauss with the interval of 500 gauss, but {\bf Stokes $I/I_c$ from 0 Gs},   and the same case under the SW umbral model atmosphere (blue solid or dotted lines) \cite{Stellmacher70,Stellmacher75}. 
\label{fig:line5324}}
\end{figure}

 Figure  \ref{fig:5324stokes}  shows the Stokes parameters $I$, $Q$, $U$, $V$ of the FeI$\lambda$5324.19\AA{} line,
calculated with the VAL-C atmospheric model  and a homogeneous magnetic field B from 500 to 4000 gauss, inclination of the field $\psi$=30$^\circ$, azimuth $\varphi$=22.5$^\circ$ and $\mu =1$. 
Figure \ref{fig:line5324} shows the comparison with the SW umbral model atmosphere [Stellmacher \& Wiehr (1970), Stellmacher \& Wiehr, (1975) \cite{Stellmacher70,Stellmacher75}].  Because we chose the azimuth angle of the field is 22.5$^\circ$, the Stokes parameter $Q$ is equal to $U$ if the magneto-optical effect is ignored. Thus, the difference between Stokes parameters $Q$ and $U$  in Figure    \ref{fig:5324stokes} and $\ref{fig:line5324}$   provides the information of the Faraday rotation of the  plane of polarization, in the real situation that that the magneto-optical effect cannot be ignored.

\begin{figure}[H]
\centerline{\includegraphics[width=70mm]{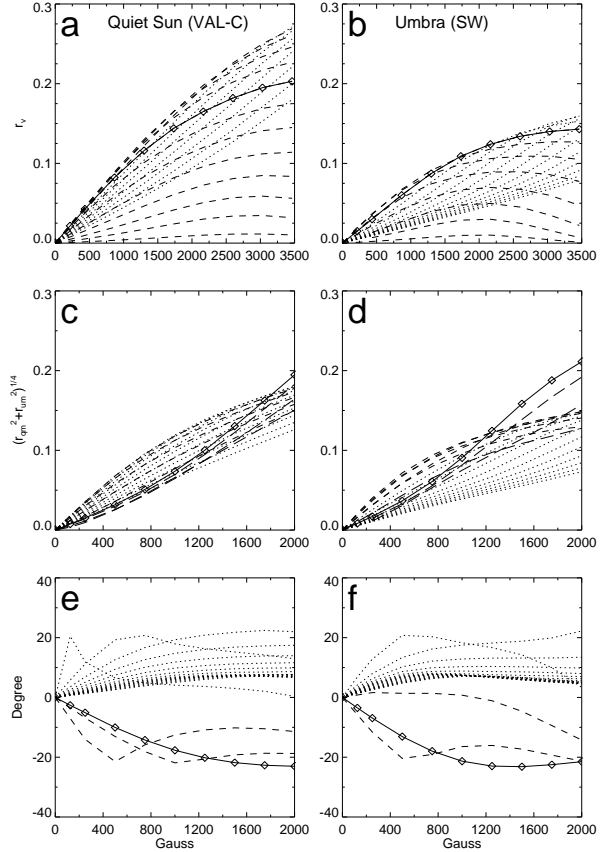}} 
\vspace{0.5cm}
\caption{The relationship amount the Stokes parameters $r_v$ {\bf and $(r^2_{qm}+r^2_{um})^{1/4}$ (i.e. $V/I_c$, $[(Q/I_c)^2+(U/I_c)^2]^{1/4})$ of} the FeI$\lambda$5324.19\AA{} line and magnetic strength in the wing from 0.005 to 0.185\AA{} with the interval of 0.01\AA{} in the quiet Sun (left) and umbral (right) model atmospheres. e-f) The ``error azimuthal angles" of the transverse field inferred due to the magneto-optical effect.  The others are the same with Figure \ref{fig:line5324}.  The $V/I_c$ relates to the longitudinal component of the field, while $Q/I_c$ and $U/I_c$  to the transverse one.
\label{fig:moline5324fra}
}\end{figure}

Figure \ref{fig:moline5324fra} shows the the Stokes parameters $Q$, $U$ and $V$ calculated in Figure \ref{fig:line5324} in a almost equivalent form.    
It  presents   the curve clusters on the  relationship between the strength of magnetic fields and the Stokes parameters $Q$, $U$ and $V$, and the ``error azimuthal angles" of the transverse field inferred due to the magneto-optical effect with different wavelengths of the FeI$\lambda$5324.19\AA{} in the quiet Sun  (VAL-C) \cite{Vernazza81} and umbral  \cite{Stellmacher70,Stellmacher75} model atmosphere. The solid lines in Figures \ref{fig:moline5324fra}a and \ref{fig:moline5324fra}b show Stokes $V/I_c$ in the wing of  FeI$\lambda$5324.19\AA{} at 0.075\AA{} from the line center, while other solid lines in Figure \ref{fig:moline5324fra}c-f  show $[(Q/I_c)^2+(U/I_c)^2]^{1/4}$ and ``error azimuthal angles"  of wavelength 0.005\AA{} from the line center (i.e. almost at the line center).
 
Figure \ref{fig:moline5324fra} shows that the quiet Sun and sunspot atmosphere have different sensitivities of Stokes parameters of the spectral line. The ratio between the Stokes parameter $r_v$ (i.e. $V/I_c$) of the quiet Sun and sunspot umbra is about 1.8, and the linear approximation to  Stokes $V$ can be used  for relatively weak magnetic field only below 2000G for the quiet Sun and 1000G for the umbra, respectively.  
Similarly, the ratio between the {\bf Stokes parameter $(r^2_{qm}+r^2_{um})^{1/4}$ } (i.e. $[(Q/I_c)^2+(U/I_c)^2]^{1/4}$) of the quiet Sun and sunspot umbra is about 1.5 normally and it changes with wavelengths. The linear approximation can be used  for relative weak magnetic field only below 1000G for the quiet Sun and 600G for the umbra, respectively.   
These 
reflect the nonlinearity between the strength of magnetic fields and the measured polarized components of the spectral line.

It is presented from the observations  \cite{Zhang00}  that the magneto-optical effect is a notable problem for the measurement of the transverse field with the FeI$\lambda$5324.19\AA{}  line. Although the magneto-optical effects have been neglected in some studies of the transverse magnetograms obtained by the Huairou magnetograph  (Wang et al. 1992) \cite{WangHM92}, its influence is more obvious near the line center than in the far wing.

The  magneto-optical effect is  a notable problem for the diagnosis of magnetic field in solar active regions using magnetic sensitive lines by vector magnetographs (Landolfi and Landi Degl'Innocenti, 1982; West and
Hagyard, 1983; Skumanich and Lites, 1987) \cite{Land82,West83,Skum87}, { because it  influences the determination} of the azimuthal angle of the transverse magnetic field. 
 The corresponding rotation of the azimuthal angles of polarized light related to the transverse components of the fields can be found in Figure \ref{fig:moline5324fra}.

\begin{figure}[H] 
\centerline{\includegraphics[width=65mm]{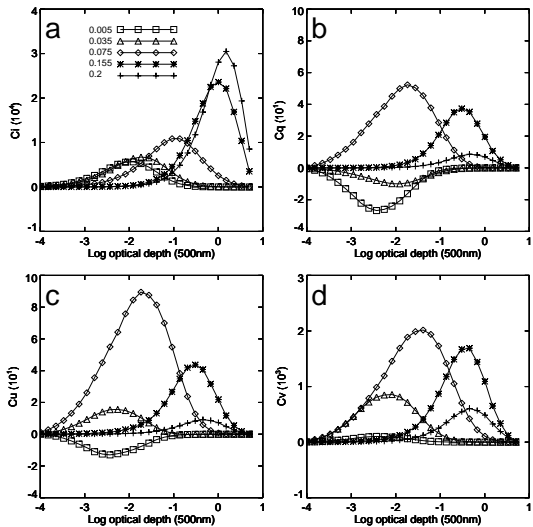}}
\vspace{1cm}
\caption{
The contribution functions  $C_i,C_q,C_u$, and $C_v$ of  Stokes parameters $I$, $Q$, $U$ and $V$ of the
FeI$\lambda$5324.19\AA{} line  are numerically calculated under a VAL-C model
atmosphere at the wavelengths $\triangle\lambda$=0.2 (cross), 0.155 (star),  0.075 (diamond), 0.035 (delta) and 0.005  (black) \AA{} from the  line center, with the magnetic field B=1000 Gauss, the inclination
$\psi$=30$^\circ$, azimuth $\varphi$=22.5$^\circ$, and $\mu =1$.
\label{fig:cbf5324}}
\end{figure}

The ``error azimuthal angle" caused by the magneto-optical effect is defined as the difference between that of the magneto-optical effect and that of the non magneto-optical effect,
\begin{equation}
\bigtriangleup\varphi=0.5\left[\textrm{tg}^{-1}\left(\frac{U_f}{Q_f}\right)-\textrm{tg}^{-1}\left(\frac{U}{Q}\right)\right],
\end{equation}
where the subscript $f$ marks that the Faraday rotation of the polarized spectral line {\bf has been considered.}  It is found that the ``error azimuthal angle" of the transverse field obviously relates to the wavelength from the line center to the wing  of the  FeI$\lambda$5324.19\AA{} line \cite{Zhang00}. It is found that the ``error azimuthal angle"  of the transverse field is about 10$^\circ$ near the line center and about 5$^\circ$ at -0.15\AA{} from the line center as the magnetic field is 1000 Gauss.  
Similar result for Stokes parameters $I$, $Q$, $U$ of the  FeI$\lambda$5324.19\AA{} line  under the cool umbra model (Stellmacher \& Wiehr, 1970) \cite{Stellmacher70} can be found also. The ``error azimuthal angle" of
the transverse field is an order of about 20$^\circ$ near the line center and about 10$^\circ$ at -0.15\AA{}. 
It is noticed that the maximum errors of the azimuthal angles of transverse components of the magnetic field can reach $\pm 20^\circ$ due to the magneto-optical effect in Figure \ref{fig:moline5324fra}. 


{ Although we display some examples of the magneto-optical effects on the FeI$\lambda$5324.19\AA{} line, but these provide a basic estimation on the order of ``error azimuthal angles" for the measurement of the transverse magnetic field by the FeI$\lambda$5324.19\AA{} line only. The ``error azimuthal angle" relates to the inclination angle and intensity of the magnetic field from the calculations.  The magneto-optical effects  with the  FeI$\lambda$5250.22\AA{} line were calculated by Solanki (1993) \cite{Solanki93} and also West \& Hagyard (1983) \cite{West83}. It seems the different sensitivity of the magneto-optical effect relative to the FeI$\lambda$5324.19\AA{} line. This is consistent with the result obtained by Landolfi \& Landi Degl’Innocenti (1982) \cite{Land82}, i.e. the error in the azimuthal angle is larger for intermediate values of the Zeeman splitting (0.5$\leq$ $v_H$$\leq$2.5), if we keep in mind that the land\'e factor of the FeI$\lambda$5250.22\AA{} line is 3.0 and that of the FeI$\lambda$5324.19\AA{} line is 1.5, and the equivalent widths of these lines are different. 
We also notice that the calculated amplitude of the magneto-optical effect relates to the choices of the parameters of solar model atmospheres and the simplification of radiative transfer equations etc.}
 
\subsection{Formation Layers of  Spectral Lines}

It is important to know where in the solar atmosphere a given
spectral line formed. This information is provided by the
contribution function $C_\textbf{  I}$ defined by Stenflo (1994) \cite{Stenflo94}
\begin{equation}
\textbf{  I}(0)=\int _0 ^{\infty}C_\textbf{  I}(\tau_c)d\tau_c=\int _{-\infty}
^{\infty}C_\textbf{  I}(x)dx
\end{equation}
for the Stokes vector $\textbf{  I}=(I,Q,U,V)$, where $C_\textbf{ 
I}(x)=(\textrm{ln}10)\tau_c C_\textbf{  I}(\tau_c)$ and $x=\textrm{log}\tau_c$,  $\tau_c$
is the continuum optical depth at 5000\AA. It provides the contribution to the emergent Stokes parameters from the different layers of the solar atmosphere. 

Now, we can define the equivalent source functions $S_\textbf{  I}^\star$
(Zhang, 1986)\cite{Zhang86} 
\begin{equation}
\begin{array}{c}
\displaystyle 
S_I^\star=S-\frac{1}{\eta _0 +\eta _I}[\eta _Q Q+\eta _U U+\eta _V V]~~~~~~\\
\displaystyle S_Q^\star=-\frac{1}{\eta _0 +\eta _I}[(\eta _Q(I-S)+\rho _V U-\rho _U V)]~\\
\displaystyle S_U^\star=-\frac{1}{\eta _0 +\eta _I}[(\eta _U(I-S)-\rho _V Q+\rho _Q V)]~\\
\displaystyle S_V^\star=-\frac{1}{\eta _0 +\eta _I}[(\eta _V(I-S)+\rho _U Q-\rho
_Q U)].\label{eq:line29}
\end{array}
\end{equation}

Similar studies have been done by Jin (1981), Song et al. (1990), Qu et al. (1997) \cite{Jin81,Song90,Qu97}.
These represent formal radiative sources of Stokes parameters  contributing to the emergent Stokes parameters at  different layers of the solar atmosphere. The Unno-Rachkovsky equations of radiative transfer of polarized light can be written in a compact form
\begin{equation}
\mu \frac{d \textbf{  I} }{d\tau_c}=(\eta _0 +\eta _I)(\textbf{  I}-S_\textbf{ 
I}^\star ).\label{eq:line30}
\end{equation}
It is the same as eq. (11.129) of \cite{Stenflo94} and $\mu$ is the cosine of the heliocenter angle.  
In the following, we analyze the line formation in the center of the solar disk, where $\mu=1$. We  notice that the contribution functions in the far wings also bring some information of the continuum, while we also notice that the continuum does not significantly contribute to the Stokes parameter $V$ of the lines near the working wavelengths of the Huairou Magnetograph to influence the estimation on the  formation layers of the Stokes parameters.  The contribution function $C_\textbf{  I}$ is
\begin{equation}
C_\textbf{  I}(\tau_{ci})=\frac{1}{\mu} S_\textbf{  I} ^\star
\exp\left[-\frac{1}{\mu}\int _0 ^{\tau _{ci}}(\eta _0 +\eta _I) d\tau_c\right]
(\eta _0 +\eta _I),
\end{equation}
where $\tau_{ci}$ is the continuum optical depth at the $i$-th depth point. We can find that this contribution function obviously relates to the simple form of the radiative transfer equation (\ref{eq:line30}) and can be used in the numerical calculation of the radiative transfer equations conveniently.  

\begin{figure*}[t]  
\begin{center}
\includegraphics[angle=0,width=140mm]{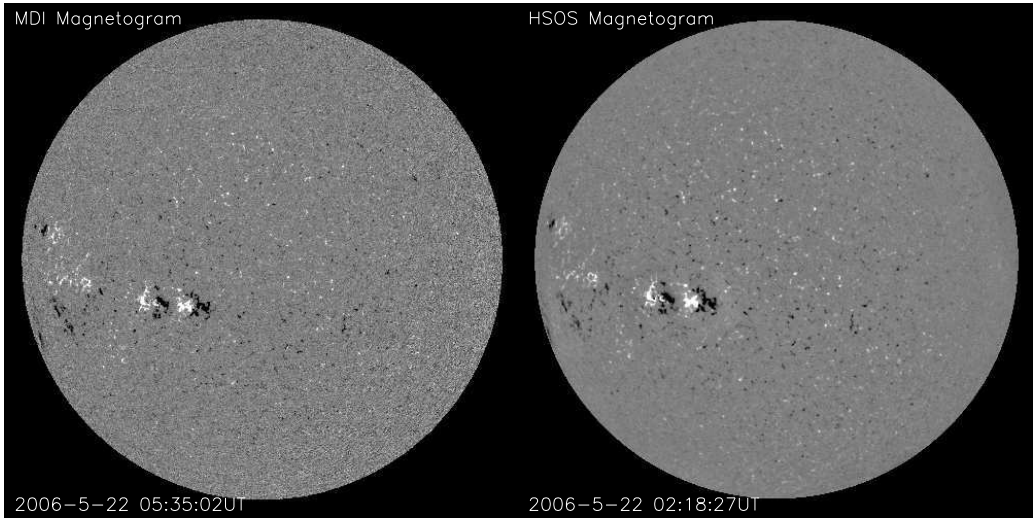} 
\end{center}
\vspace{-7.0cm}
\caption{The comparison between the full disk longitudinal magnetograms observed form MDI of SOHO satellite (left) and vector magnetograph at Huairou (right) on 2006 May 22.  From \cite{Zhang07} \label{fig:tele5} }
\end{figure*}

Figure $\ref{fig:cbf5324}$ shows the formation depths of the FeI$\lambda$5324.19\AA{} line in the solar VAL-C model atmosphere with different wavelengths within 0.005\AA{} - 0.2\AA{} from the line center. In the far wing $\bigtriangleup\lambda=0.2$\AA{}, Stokes parameters $I$, $Q$, $U$, $V$ mainly form near the relative deep solar photosphere $\textrm{log}\tau_5\sim (-1,0)$, while near the line center $\bigtriangleup\lambda=0.005$\AA{} Stokes {\bf parameters mainly form} in the relative high photosphere $\textrm{log}\tau_5\sim (-3,-1)$, where $\tau_5$ means the optical depth of continuous spectrum at 5000\AA. 

The calculation provides that the Stokes parameters of the FeI$\lambda$5324.19\AA{} line form below the solar temperature minimum region  (about $\tau_{500\textrm{nm}}=10^{-4}\!\sim\! 10^{-3}$) in the model atmosphere (Vernazza et al., 1976) \cite{Vernazza76}.  
This means that FeI$\lambda$5324.19\AA{} is a typical photospheric line. { The formation layer of the spectral  lines in sunspots is different from that in the quiet Sun.}

\subsection{ Measurements of Full Solar Disk Magnetic Field}

Solar Magnetism and Activity Telescope (SMAT) is operated at Huairou Solar Observing Station, National astronomical Observatories of China started at 2003 (Zhang et al., 2007) \cite{Zhang07}.
A birefringent filter is centered at 5324.19\AA{} for the measurement of vector magnetic field and  the bandpass of the filter is 0.1\AA{}. 
Figure \ref{fig:tele5}   shows the comparison between both longitudinal magnetograms obtained by MDI of SOHO satellite and SMAT on 2006 May 22. It can be found the basic morphological correlation between both magnetograms.  Some slight differences of both magnetograms  are probably caused by the different seeing condition, observing noise,  and also the data reducing methods.

\begin{figure*}[t]  
\vspace{1.0cm} 
\begin{center}
\includegraphics[angle=0,width=100mm]{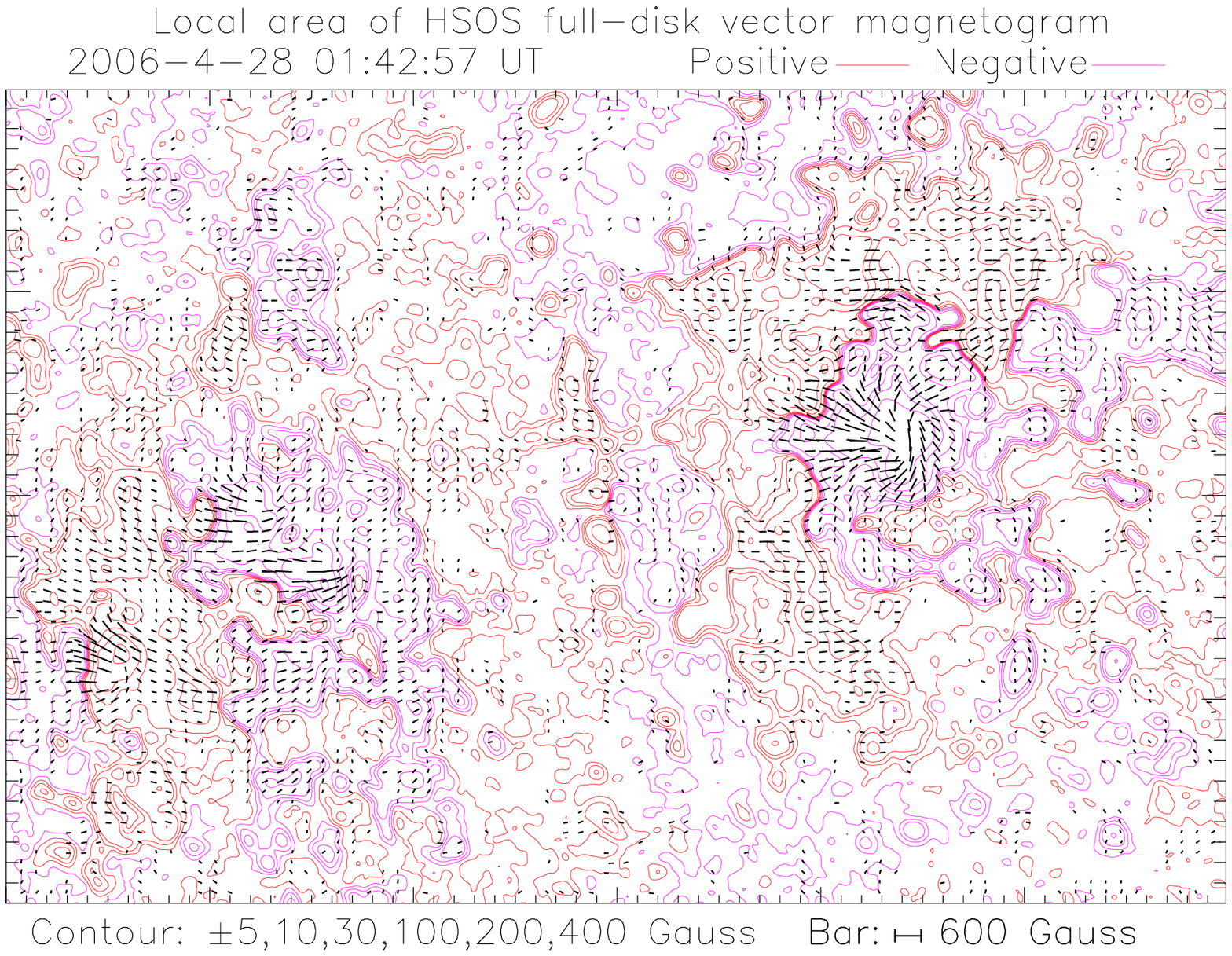} 
\end{center}
\caption{Local area of full disk vector magnetogram on 2006 April 28. The bars mark the transverse magnetic field.  From \cite{Zhang07} \label{fig:tele7}}
\end{figure*}

\begin{figure}[H]  
\begin{center}
\includegraphics[angle=0,width=80mm]{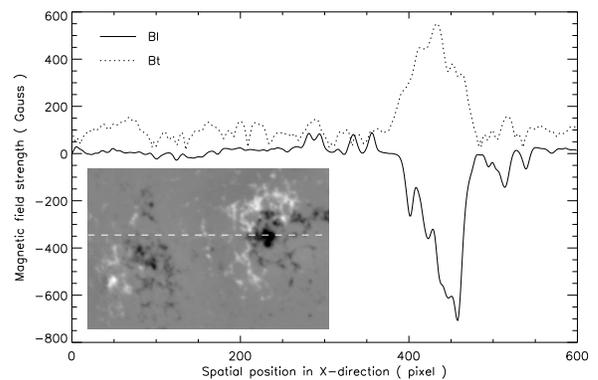}
\end{center}
\caption{The intensity distribution of longitudinal and transverse magnetic field along the dashed beeline in the local area of full disk magnetogram on 2006 April 28 (bottom left), which vector magnetogram is shown in Figure \ref{fig:tele7}.  From \cite{Zhang07} 
\label{fig:tele8}}
\end{figure}

A local area of the full disk vector magnetogram is shown in Figure \ref{fig:tele7}. The strong transverse components of the field  extend from the strong longitudinal field in the active regions. It is estimated that in this region the sensitivity of the longitudinal component of magnetic field is about or lesser than the order of 5 gauss and the transverse one is about 100 gauss in Figure \ref{fig:tele8}. 

\begin{figure*}[t]  
\begin{center}\includegraphics[angle=0,width=140mm]{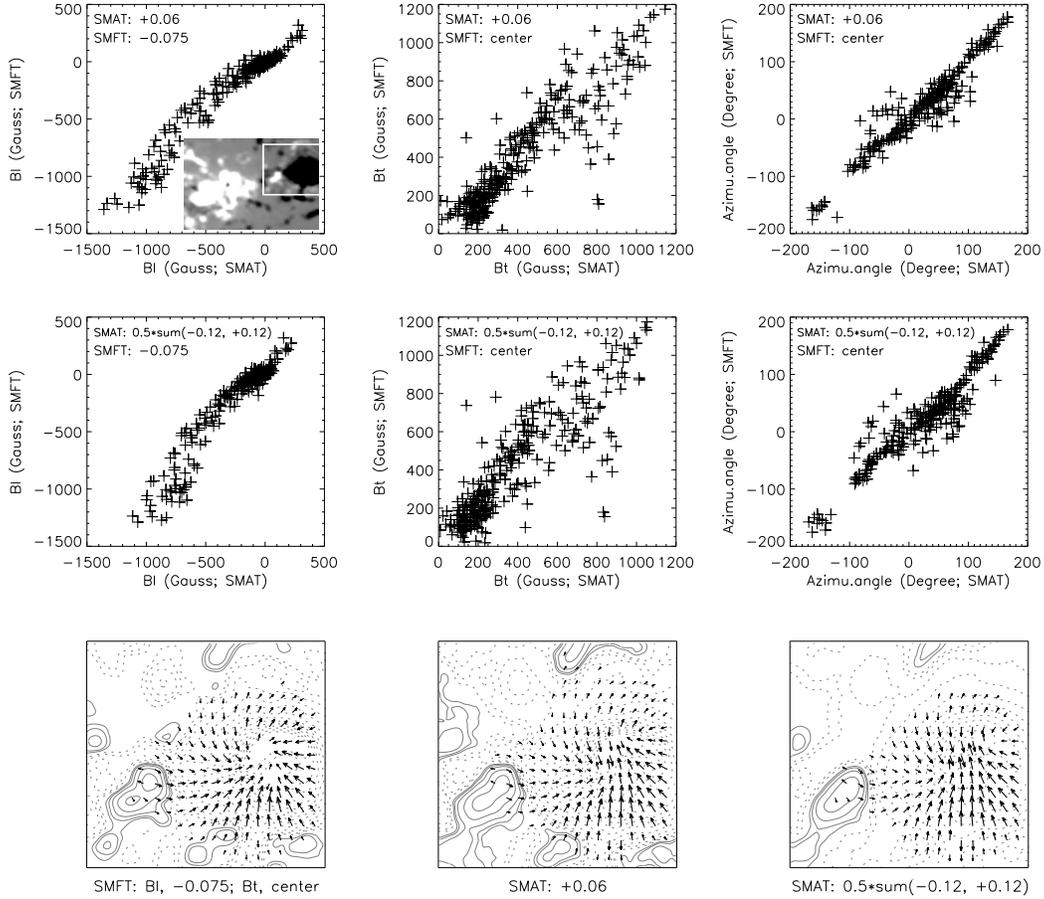}\end{center}
\vspace{-0.5cm}\caption{The relationship between the vector magnetograms observed by the SMFT and the SMAT in the region marked with the white box in {\bf the Stokes V image} at the bottom-right corner of the first panel. The relationships of the longitudinal components, the transverse components and the azimuthal angles of the vector magnetic fields of both instruments are in the first two rows, the vector magnetograms in the last row. The wavelengths of both instruments are marked in each panel. In the scatter correlation plots, a 2.5 factor has been applied to the circularly polarized signal and a 3.0 factor to the linearly polarized signals of the SMAT. From \cite{Su07}}
\label{fig:FUCO9}
\end{figure*}

Figure \ref{fig:FUCO9} shows the correlation between the vector magnetograms observed by SMFT and SMAT in an active region (Su \& Zhang, 2007) \cite{Su07}. The relationships of the longitudinal components, the transverse components, and the azimuthal angles of both vector magnetic fields are shown in the figure. 
It is found a relative high correlation between both vector magnetogams observed from SMFT and SMAT.  The study on the spatial integration scanning spectra of filter magnetograph is also important for the calibration of the full-disk vector magnetic fields, which was made by Wang et al. (2010) \cite{WangX10}.

\begin{figure*}[t] 
\centerline{\includegraphics[width=85mm]{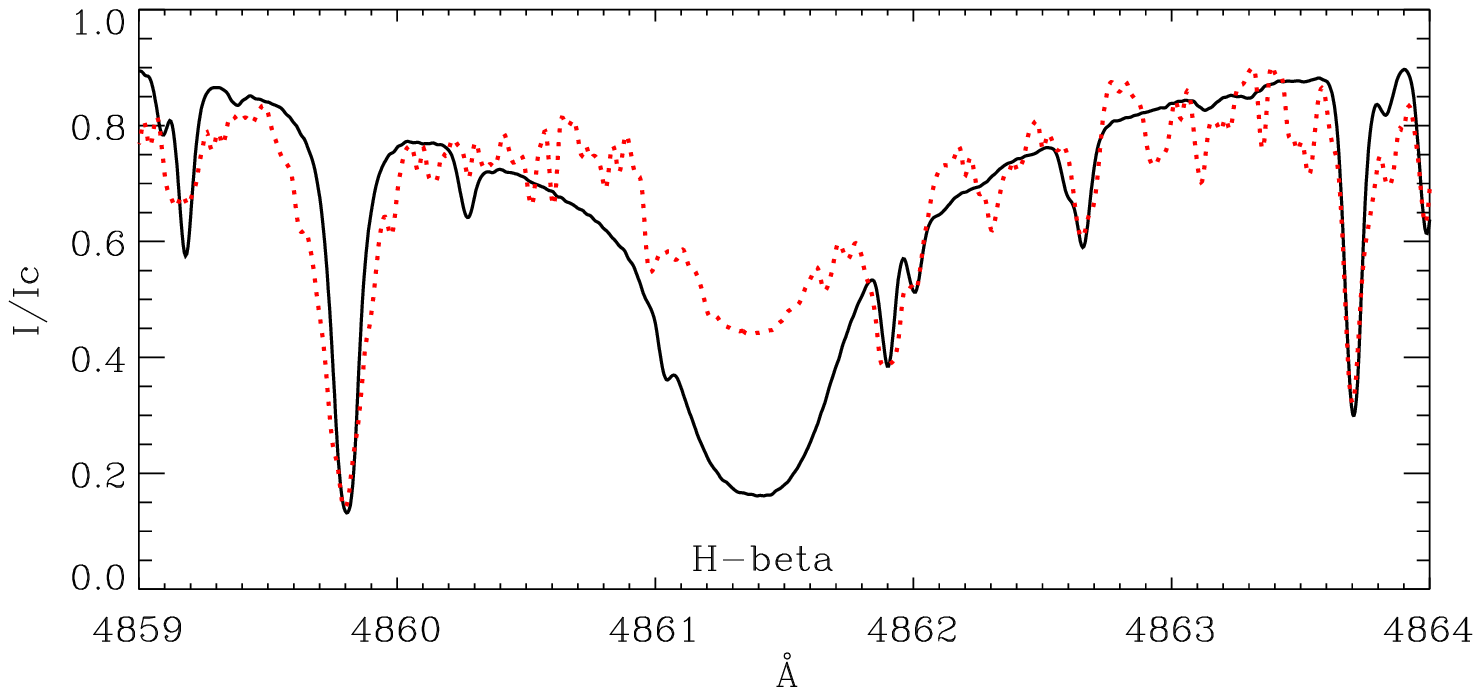}\includegraphics[width=85mm]{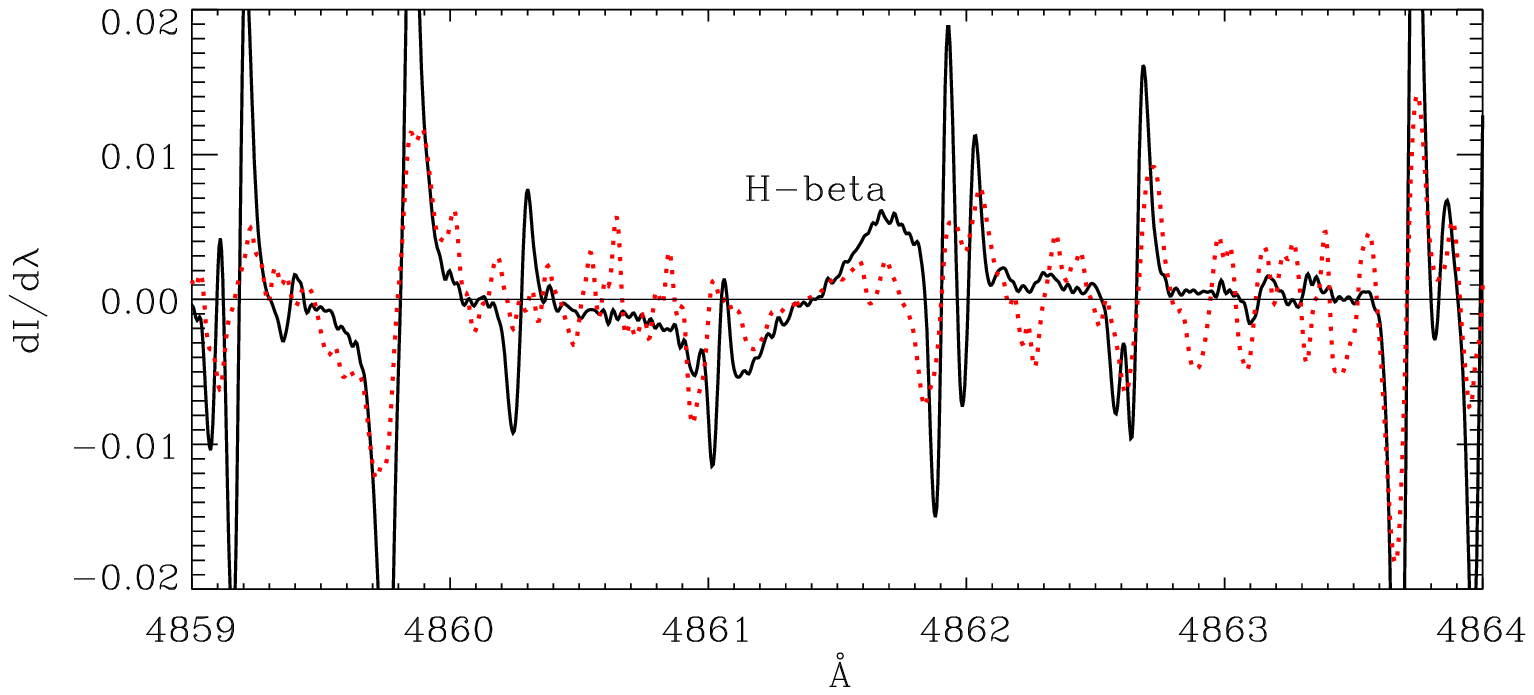}}
\caption{ {\bf Left: }H$\beta$ lines in the quiet Sun (solid line) and sunspot umbra  (dotted line)  observed by the National Solar Observatory of USA (L. Wallace, K. Hinkle, and W. C. Livingston, http://diglib.nso.edu/ftp.html). {\bf Right:} $dI/d\lambda$ inferred by the corresponding  H$\beta$ lines in above. }\label{fig:StokesHV2}
\end{figure*}

\section{Formation of H$\beta$ Line in Solar Chromospheric
Magnetic Field}

It is believed that solar active phenomena often relate to the complex configurations of coronal and chromospheric magnetic fields. The observations of chromospheric magnetograms were made, such as,  at the Crimean, Kitt Peak and Huairou Observatories (cf. Severny \& Bumba (1958), Tsap (1971), Giovanelli (1980), Zhang et al. (1991) \cite{SeBu8,Tsap71,Giovanelli80,Zhangetal91}). The H$\beta$ line is a working line of SMFT at  Huairou Solar Observing Station of National  Astronomical Observatories of China for the measurements of chromospheric magnetic fields (Zhang \& Ai, 1987) \cite{ZhangA87}.

\begin{table}[H] 
\caption{Parameters of H$_\beta$ line. $\gamma_R$ is the damping parameter, $gf$ is the oscillator strength and  $gf_w$ is the normalized oscillator strength. \cite{Bethe57,ZhangA87}}
\begin{center}
\begin{tabular}{ccccc}
\hline  {Transition} & {WL (\AA)} & {$\gamma_R\times10^8$} & {$gf$} & {$gf_w$} \\ 
\hline
$2p_{1/2} - 4d_{3/2}$   & 4861.279   &  6.5452  &  0.2436 &  0.2552        \\
$2s_{1/2} - 4p_{3/2}$   & 4861.287   &  0.8328  &  0.13702 &  0.1435        \\
$2p_{1/2} - 4s_{1/2}$   & 4861.289   &  6.3122  &  0.00609 &  0.0068        \\
$2s_{1/2} - 4p_{1/2}$   & 4861.298   &  0.8328  &  0.0685 &  0.0718        \\
$2p_{3/2} - 4d_{5/2}$   & 4861.362   &  6.5452  &  0.4385 &  0.45935        \\
$2p_{3/2} - 4d_{3/2}$   & 4861.365   &  6.5452  &  0.0487 &  0.0510        \\
$2p_{3/2} - 4s_{1/2}$   & 4861.375   &  6.3122  &  0.0122 &  0.01278        \\
\end{tabular}\end{center}
\label{tab:Hbetaline}
\end{table}

Figure \ref{fig:StokesHV2} shows the solar H$\beta$ line profile  and some photospheric lines overlap in the wing. The wavelength of the H$\beta$ line is 4861.34\AA{} and its equivalent width is 4.2\AA{}. The core of the  line is formed at a height of about 1900 km (Allen, 1973) \cite{Allen73}, while different formation heights of the H$\beta$ line were proposed by various authors. Its  oscillator strength is 0.1193 and the  residual intensity at the core is 0.128 (Grossmann-Doerth \& Uexkull, 1975) \cite{Grossmann-Doerth75}. It is normally believed that the Doppler broadening forms in the  core of the H$\beta$ line, and  resonance damping and Stark broadening in the wing. The H$\beta$ line is composed of 7 lines and all located within a width of nearly 0.1\AA{}. We can use the formulae given by Bethe \& Salpeter (1957) \cite{Bethe57} to calculate  individual component lines, the wavelength shift,  the normalized oscillator strength and damping constant, which are shown in Table \ref{tab:Hbetaline}. In comparison with the results given by Allen (1973) \cite{Allen73} and Garcia \& Mack(1965) \cite{Garicia65}, these available parameters showed slight difference in the mean wavelengths. When the magnetic field alone is present, the H$\beta$ line shows the anomalous Zeeman effect. In the solar atmosphere, both magnetic field and interatomic microscopic electric field are present, { the wave functions of the hydrogen atom energy levels become degenerate,  and} different wave functions relate to complicated energy shifts.

\subsection{Radiative Transfer of H$\beta$ Line}

We  study the formation of polarized light of the H$\beta$ Line with   Eq. (\ref{radtrf}), Unno-Rachkovsky equations of polarized radiative transfer,  in the solar atmospheric magnetic field. The {\bf source function  $S_\nu$} in Eq.  (\ref{eq:souf2}) for the line spectrum is departure from LTE with coefficients $b_l$ and $b_u$ in Eq. (\ref{eq:souf1}).  We also assume the continuum to be LTE. 

As a magnetic field is present, the broadening of the hydrogen line, such as H$\beta$, should be the joint effect of the magnetic field and the microscopic electric field distributed according to the Holtsmark statistics. 
It is assumed that the microscopic electric field is isotropic in the solar atmosphere, and their contribution to the
polarization of the emergent light of the H$\beta$ line is negligible, thus the angular dependence and polarization properties of the $\pi$ and $\sigma$ components of the transitions of the H$\beta$ line can be obtained from the classical theory.

In comparison with formulae of the non-polarized hydrogen line proposed by Zelenka (1975) \cite{Zelenka75}, { Zhang \& Ai (1987) \cite{ZhangA87} introduced the  absorption coefficients of H$\beta$ line in the magnetic field atmosphere in the form}
\begin{eqnarray}
\label{eq:HB2}
\eta_{p,l,r}&=&\eta_{0}{\sum_if_{wi}H(a_i,v_i(\triangle m))\int_0^{\beta_l} W(\beta,r_0}/D)d\beta\nonumber\\
&+&\frac{1}{4\pi}\eta_{0}{\sum_{j}f_{wj}\int_{0}^{2\pi} \int_{0}^{\pi}\int_{\beta_l}^{\infty} H\left(S(a_j, \lambda)\right)}\\
&&W(\beta,r_{0}/D)d\beta\sin\theta d\theta d\chi,\nonumber
\end{eqnarray}
and
\begin{eqnarray}
\label{eq:HB3}
\rho_{p,l,r}&=&2\eta_{0}{\sum_if_{wi}F(a_i,v_i(\triangle m))\int_0^{\beta_l} W(\beta,r_0/D)d\beta}\nonumber \\
&+&\frac{1}{2\pi}\eta_{0}{\sum_{j}f_{wj}\int_{0}^{2\pi}\int_{0}^{\pi}\int_{\beta_l}^{\infty}F\left(S(a_j, \lambda)\right)}\\
&&W(\beta,r_{0}/D)d\beta\sin\theta d\theta d\chi,\nonumber
\end{eqnarray}
where $S(a_j, \lambda)=\left(a_{j},\frac{\triangle\lambda-\triangle\lambda_j(\theta,\chi,\beta,\triangle
m)} {\triangle\lambda_{D}}\right)$ and
$f_{wi}$ and $f_{wj}$ are normalized oscillator strengths of polarized subcomponents of the H$\beta$ line. $W(\beta,r_{0}/D)$ is the Holtsmark distribution function of the microscopic electric field. $\beta _l$ is a criterion to measure the relative size of the shift caused by the electric field, relative to the fine structure of the hydrogen line. If $\beta _l$ is very large, then the effect of the electric field can be negligible.

For H$\beta$ line, we can choose $\beta_l=0.1189/F_0=9.512\times 10^{12}N_e^{2/3}$. Then,  the second terms on the right of formulae (\ref{eq:HB2}) and (\ref{eq:HB3}) for the lower solar atmosphere are  the main terms and the effect of the interatomic electric field is all important. As we go up in the chromosphere, the density of charged ions falls and the first term becomes more and more important.

The first terms in the formulae (\ref{eq:HB2}) and (\ref{eq:HB3}) reflect the contribution under the weak microscopic electric field, thus the approximation of Zeeman splitting is suitable. There $v_i(\triangle m=1)$,
$v_i(\triangle m=0)$ and  $v_i(\triangle m=\!-1)$ connected  with Zeeman splitting are following, for $k=1,2,3$,
$\displaystyle v_i(\triangle m=2-k)=v-v_m^{(k)},$  and
\begin{equation}
\label{eq:HB8}
\displaystyle v_m^{(k)}=\frac{e\lambda^2H}{4\pi m_e^2\triangle\lambda_D} [(G-G')M-G'(2-k)].
\end{equation}
Here, $M$ is the magnetic quantum number of the lower energy level, $G$ and $G'$ are the land\'e factors for the lower and upper levels. $f_{wi}$ is the normalized oscillator strengths of the Zeeman sub-components of the H$\beta$  line. 

While the second terms in the formulae (\ref{eq:HB2}) and (\ref{eq:HB3}) reflect the common contribution of the magnetic field and strong microscopic electric field to the line
broadening. $\triangle\lambda_j$ is the shift of the polarized $j$-component of the spectral line caused by the magnetic and strong statistical electric field. $f_{wj}$ is the normalized
oscillator strength of the polarized sub-components of the H$\beta$ line in the magnetic and microscopic electric field.  

The damping parameter of $i$-th  component of the line is
$\displaystyle a_i=\frac{\gamma_i\lambda_0^2}{4\pi c\triangle\lambda_D},$ 
where
$\displaystyle \gamma_i=(\gamma_{\textrm{radiation}}+\gamma_{\textrm{resonance}}+\gamma_{\textrm{electron}})_i.$ 

Now, a hydrogen atom and a perturbing, charged particle have been considered in the external magnetic and electric field. The hamiltonian of the hydrogen atom can be written
\begin{equation}
\hat{H} =\hat{H}_0+\hat{V}_H+\hat{V}_F,\label{eq:HB11}
\end{equation}
where the first term on the right is the unperturbed hamiltonian, the second and third terms represent the perturbation of the magnetic field and electric field, respectively. The  perturbation equation is
\begin{equation}
(\hat{H}_0+\hat{V}_H+\hat{V}_F)\mid\psi \rangle=(E_0+E' )\mid\psi \rangle,\label{eq:HB12}
\end{equation}
where $E_0$ and $E'$ are the energy eigen value and perturbation, respectively. One can calculate the shifts $\triangle\lambda_j$ of the sub-components of the spectral line by means of Eq. (\ref{eq:HB12}). Normally, the general solution of the eigenvalue problem for the hydrogen lines does not depend on the direction of the magnetic field and microscopic electric field. The choice of the coordinate system for the calculation of the Hermitian matrix of the perturbing Hamiltonian was made by Casini \& Landi Degl’Innocent (1993) 
 \cite{Casini93}.

\subsection{Numerical Calculation of  H$\beta$ Line}

\begin{figure}[H] 
\centerline{\includegraphics[width=70mm]{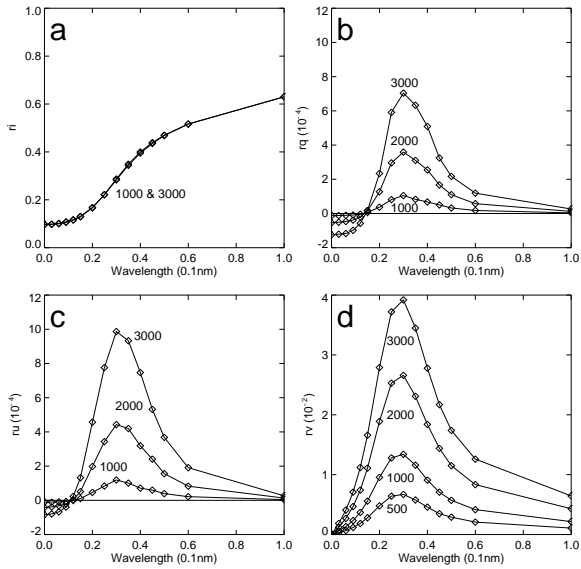}}
\vspace{0.3cm}
\caption{Profiles of Stokes parameters $r_i,r_q,r_u$, and $r_v$ (i.e. $I/I_c$, $Q/I_c$, $U/I_c$  and $V/I_c$) of the H$\beta$ line calculated for  the VAL model atmosphere, a homogeneous magnetic field intensity 1000-3000 gauss of inclination $\psi$=30$^\circ$, azimuth $\varphi$=22.5$^\circ$ and $\mu =1$. $I_c$ is the continuum.} \label{fig:HB2}
\end{figure}

For analyzing the general properties on the formation of the H$\beta$ line, we used the VAL mean quiet chromospheric model $C$ (Vernazza et al., 1981) \cite{Vernazza81}  and relevant non-LTE departure coefficients. We  assumed that, when the degeneracy of energy levels disappears under the action of the magnetic field and microscopic electric field, each magneton energy level keeps its original departure coefficient. It is only an approximation.
The emergent Stokes profiles of  H$\beta$ line are shown in Figure \ref{fig:HB2}. 
The calculation results show that the influence of the magneto-optical effect for Stokes parameter $V$ is insignificant, but that for Stokes parameters $Q$ and $U$  in the H$\beta$ line center is significant, for example the error angle of the transverse field is about 7$^\circ$. { Even though， this means that the influence of the magneto-optical effect for H$\beta$ line is weak than that of the photospheric lines, such as FeI$\lambda$5324.19\AA{} line. }       

By comparing the calculated signals of H$\beta$ line for the measurement of chromospheric magnetic field in Figure \ref{fig:HB2}  with the  photospheric one with the FeI$\lambda$5324.19\AA{} line   in Figure \ref{fig:line5324}, it is found that their ratio on the peak values of Stokes $V$ in the wing of both lines is about a order of 0.18 (0.04/0.22), while the that of Stokes $Q$ is 0.06 (0.0007/0.012) for 3000 gauss magnetic field. Because the the core of the H$\beta$ line is too flat, this is consistent with the calculated result of the weak signals of Stokes $Q$ and $U$ about a order of $10^{-4}$ in Figure \ref{fig:HB2}. This implies that the longitudinal components of the   H$\beta$ chromospheric magnetic field are detectable, while  the challenge with the high accuracy diagnostic occur in the measurements of the transverse components of the chromospheric fields. 

\begin{figure}[H]  
\centerline{\includegraphics[width=70mm]{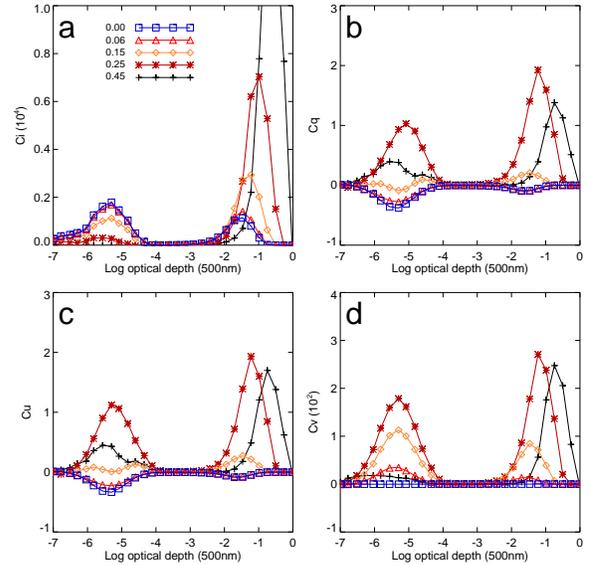}} 
\vspace{1.0cm}
\caption{Contribution functions $C_i,C_q,C_u$, and $C_v$ of Stokes parameters $I$, $Q$, $U$ and
$V$ of   the numerical solution of H$\beta\lambda$4861.34\AA{} line calculated for the VAL model atmosphere at the wavelengths $\triangle\lambda$=0.45 (cross), 0.25 (star),  0.15 (diamond), 0.06 (delta) and 0.0  (black) \AA{} from the H$\beta$ line center. B=1000 gauss, $\psi$=30$^\circ$, azimuth $\varphi$=22.5$^\circ$ and $\mu =1$. $\tau_c$ is continuum optical depth at 5000\AA{}. The horizontal coordinate is in the logarithmic scale.} \label{fig:HB5}
\end{figure}

The contribution functions of Stokes parameters of the H$\beta$ line
calculated with the VAL model atmosphere \cite{Vernazza81} 
are shown in Figure  \ref{fig:HB5}. We can see that  
in our calculation the emergent Stokes parameters at H$\beta$ line center almost form in the higher atmosphere (1500-1600 km), but that in the wing, 
for example, at -0.45\AA{} away from the H$\beta$ line center the emergent Stokes parameters reflects the information of the photospheric field  (300 km).  The formation height of the Balmer lines in the solar atmosphere excluding the magnetic field was analyzed also by several authors cf. Gibson (1973)  \cite{Gibson73}. From the calculated results, we can see that the formation heights of Stokes parameters $Q$, $U$ and $V$ are  almost the same to $I$.

It is needed to point out that real formation layers of Stokes parameters of the H$\beta$ line are more complex than theoretical cases. Numerical results of the equations of radiative transfer depend on the selection of atmospheric models and  parameters of spectral lines. For example, the formation height near the line center obviously depends on the selection of the value of the absorption coefficient of the line; and in the wing of the line that also depends on the amplitude of the line broadening. The formation layers of different kind of structures
observed in the H$\beta$ images in the solar atmosphere may be different, such as sunspot umbrae, penumbrae and dark filaments, even though these features are observed at the same wavelength in the wing of the chromospheric line. However, the computation of the  formation layers of the line provides us a reference to estimate  the possible spatial distribution of the detected magnetic field.

\subsection{Possibility of Reversal Features in H$\beta$ Chromospheric 
Magnetograms}

\begin{figure}[H]  
\begin{center}
\includegraphics[width=70mm]{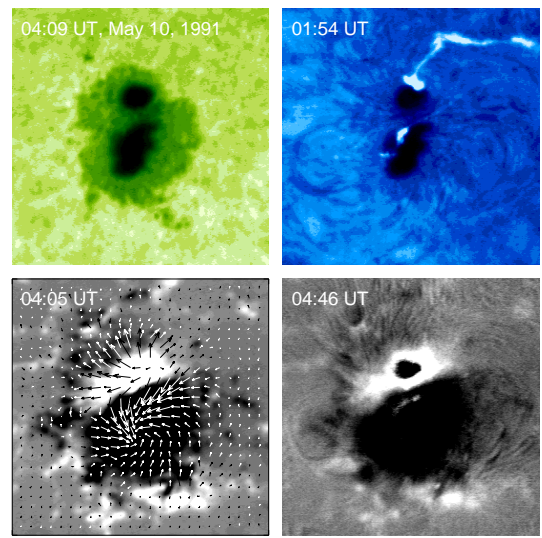}  
\end{center}
\caption{Active region NOAA 6619 on May 10 1991. Top left: Photospheric filtergram. Bottom left: Photospheric vector magnetogram. Top right: H$\beta$ filtergram. Bottom right: H$\beta$ longitudinal magnetogram. White (black) is positive (negative) polarity in the magnetograms. The size of each figure is $2.'8\times 2.'8$. 
\label{fig:ar6619}}
\end{figure}

Figure \ref{fig:ar6619} shows active region NOAA6619  on May 10 1991. The 180$^\circ$ ambiguity of transverse components of the vector magnetic field probably is an obstinate question in the measurements of magnetic fields with the polarized spectral lines.   
In Figure \ref{fig:ar6619}, the 180$^\circ$ ambiguity of transverse components of the photospheric vector magnetic field has been resolved under the assumption of the minimizing the vertical gradient of the magnetic pressure and minimizing some approximation to the field's divergence etc. cf. also Metcalf et al. (2006)  \cite{Metcalf06}. 

It is found the twisted photospheric magnetic field  formed in the compacted delta active region, and the vector magnetic field extends form photosphere into the chromosphere in the fibril-like features in the active region. The chromospheric reversal signal structure in the center of sunspot relative to positive polarity in the photosphere can be found.
The similar results have been presented, such as,  by Chen et al. (1989), Zhang et al. (1991), {\bf Almeida (1997), }Zhang (2006) \cite{Chen89,Zhangetal91,Almeida97,Zhang06a}. 
 
Due to the disturbance of the photospheric blended lines, such as  FeI$\lambda$4860.98\AA{}, nearby the H$\beta$ line center, the reversal signal in the chromospheric magnetograms relative to the photospheric ones occurs in the sunspot umbrae in Figure \ref{fig:ar6619}. It is the same with the reversal of Stokes $V$ nearby the line center in Figure \ref{fig:StokesHV2}. In the approximation of the week magnetic field (Eqs. (\ref{eq:weakfv}) and (\ref{eq:weekfild1})), the sensitivity of longitudinal  magnetic field is proportional to the variation of the line profile  $V\!\sim dI/d\lambda$ with the wavelength (Stenflo et al., 1984) \cite{Stenflo84}. 

\begin{figure}[H] 
\centerline{\includegraphics[width=60mm]{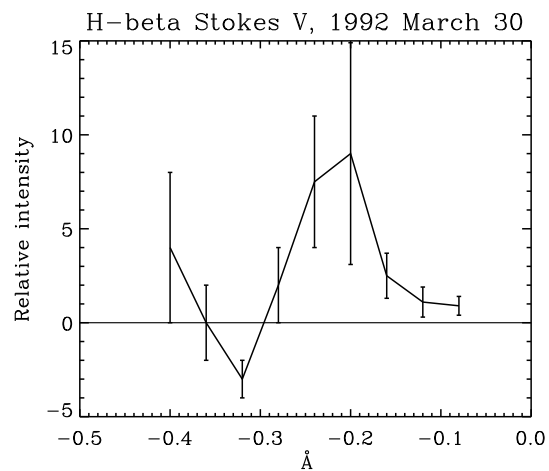}}
\vspace{0.5cm}
\caption{Statistical distribution of Stokes parameter $V$ in the blue wing of H$\beta$   line   in the umbrae of a solar active region. From \cite{Zhang93}
\label{fig:StokesHV1}
}
\end{figure}

{ Figure \ref{fig:StokesHV1} shows the statistical distribution of Stokes parameter $V$  at the different wavelengths in the blue wing from the H$\beta$ line center \cite{Zhang93} at the umbra of a solar active region. It is based on a series of the longitudinal magnetograms  observed  at Huairou Solar Observing Station.  It confirms the disturbance of FeI 4860.98\AA{} in the measurements of chromospheric magnetic fields.
}

It is noticed that in the quiet, plage regions, and even penumbrae, the influence of the photospheric blended FeI$\lambda$4860.98\AA{} line is generally insignificant. As regards the  H$\beta$ chromospheric magnetograms, we can select the working wavelength between -0.20 and -0.24\AA{}  from the line core of H$\beta$ to avoid the wavelengths of the photospheric blended lines in the wing of H$\beta$. 
A similar evidence on the disturbance of blended lines in the wing of H$\alpha$ is presented by  Hanaoka (2005) \cite{Hanaoka05}. 
 
A comparison between photospheric and chromospheric H$\beta$ magnetograms in the quiet-Sun was provided by Zhang \& Zhang(2000) \cite{ZhangM00}, who found the similar patterns of the magnetic elements in the photosphere and chromosphere. It probably brings a message that the magnetic field extends in the form of the fibril-like configuration in the quiet Sun. The  fibril-like magnetic features can also be found around the sunspots in the chromospheric magnetogram of Figure \ref{fig:ar6619}.

From the presentation above, it is found the complexity and difficulty on the diagnostic of chromospheric magnetic fields relative to the photospheric ones.

\section{Magnetic fields  in solar active regions}

\subsection{Non-potentiality of magnetic fields from observations}

\begin{figure*}[t]
\begin{center}
\includegraphics[width=85mm]{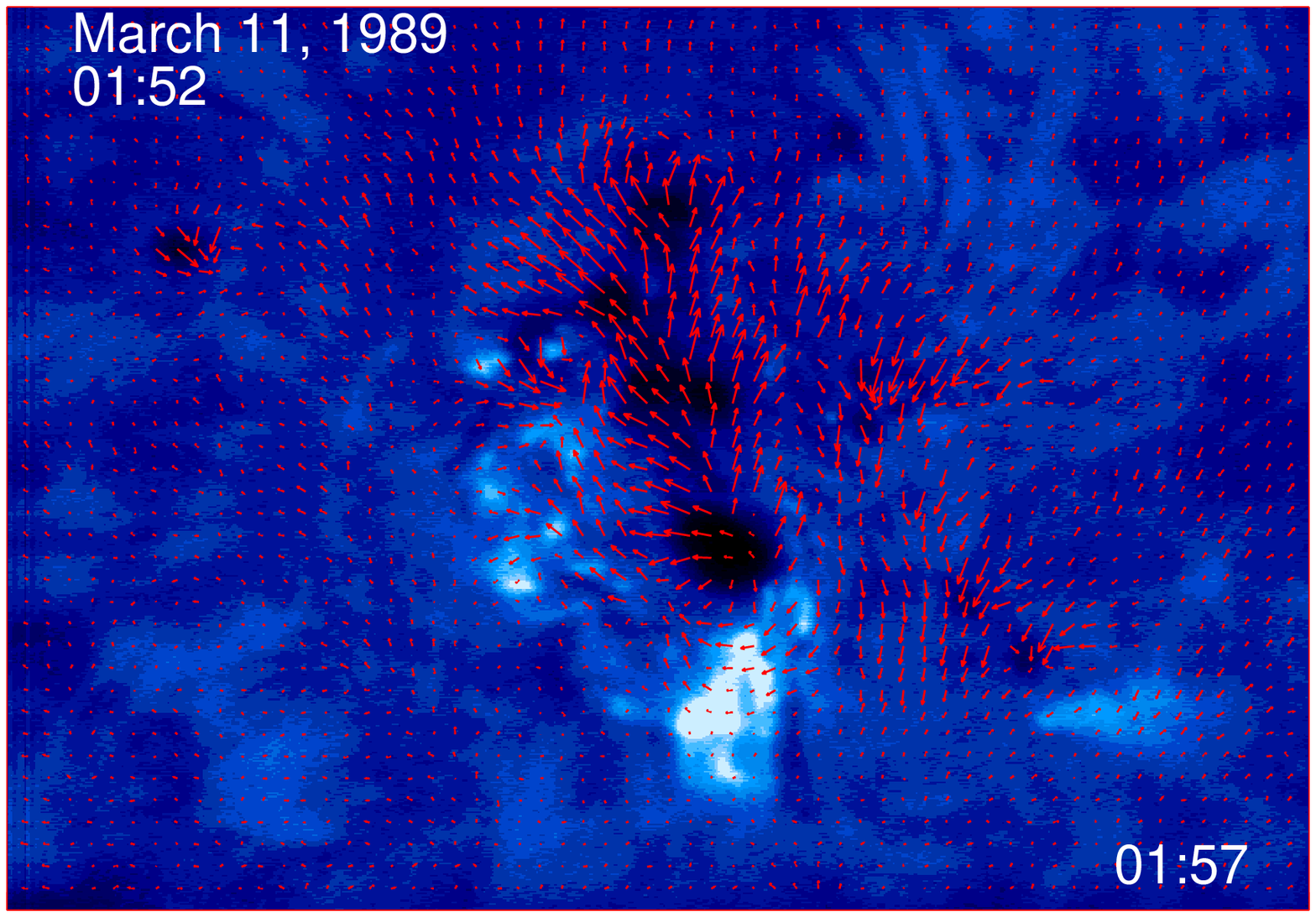} 
\hspace{0.2cm}
\includegraphics[width=85mm]{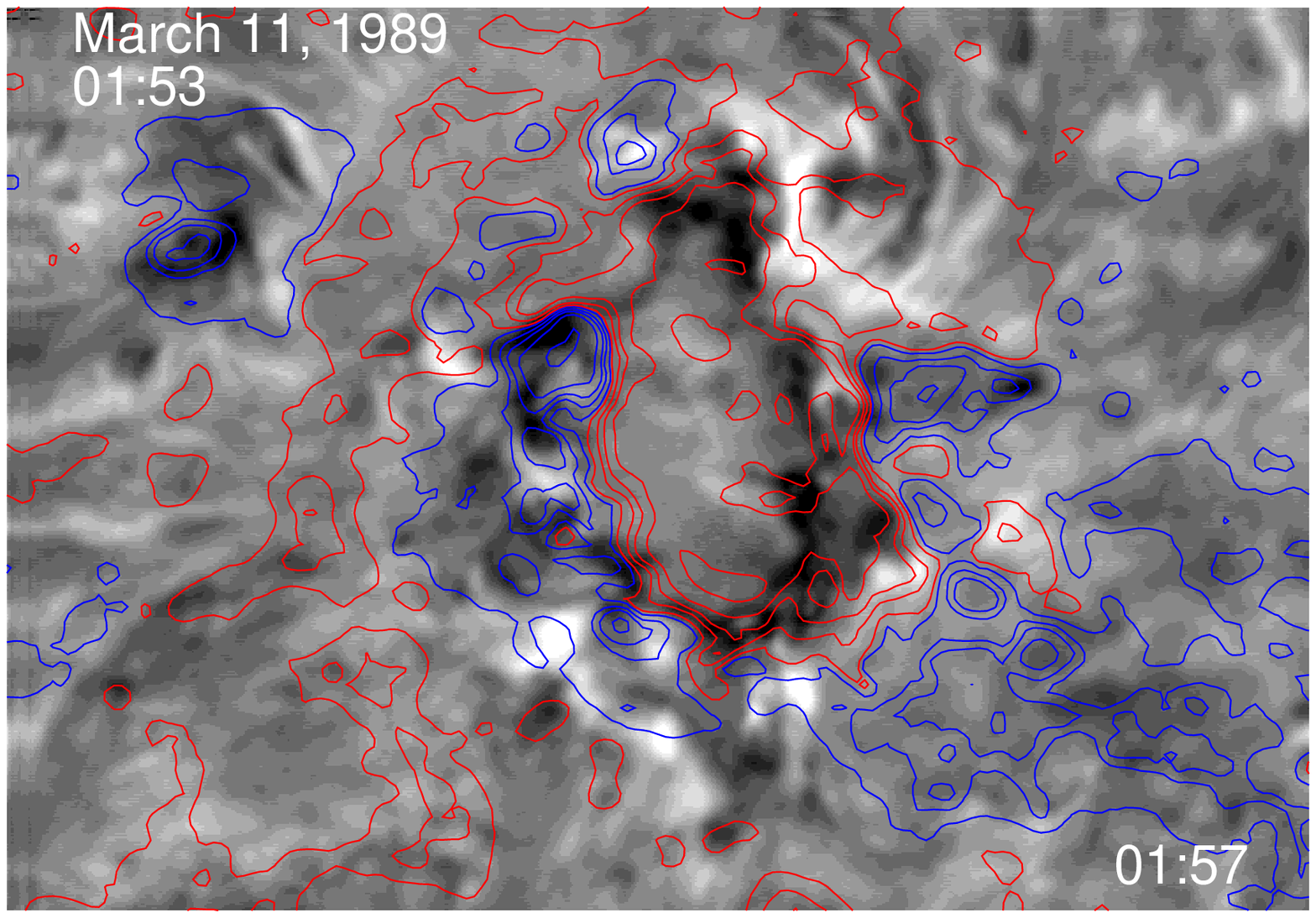} 
\end{center}
\caption{Left: A  resolved the 180$^\circ$-ambiguity of transverse components of photospheric vector magnetic field in the active region (NOAA 5395) at 01:57 UT on March 11, 1989, which is overlaid by a H$\beta$ filtergram at 01:52 UT. The arrows mark the directions of transverse magnetic field.  Right: A corresponding H$\beta$ Dopplergram of the active region at 01:53 UT with longitudinal components of magnetic field. The red (blue) contours correspond to positive (negative) fields of ${\pm}$50, 200, 500, 1000, 1800, 3000 gauss. The white (black) indicates up-(down-)ward follow. The size of magnetogram is $5.'23\times 3.'63$. The north is top and the east is at left.
\label{fig:MGsh00}}
\end{figure*}

Figure \ref{fig:MGsh00} shows the vector magnetogram and Dopplergram of active region NOAA5395 on March 11 1989, observed at the Huairou Solar Observing Station of the National Astronomical Observatories of China. 

Ai et al. (1991) \cite{Ai91} indicated that the flares   in active region NOAA5395   tend to occur on the redshift side of the inversion lines of the H$\beta$ Doppler velocity fields observed half to two hours before. A similar case can be confirmed in Figure \ref{fig:MGsh00}, although  in this Dopplergram the disturbance from the flare sites cannot be neglected, due  to the almost same observing times  between this Dopplergram and the flare.

Some evidence in this powerful flare-producing region also can be found (Zhang (1995) \cite{Zhang95c}): 1) The twisted transverse magnetic field occurred in bay-like magnetic main poles of opposite polarity in the active region. The main poles of opposite polarity in this region moved apart at a speed of about 0.15km/s. 2) The newly-sheared vector magnetic structures formed near the magnetic neutral line between the collided magnetic main poles of opposite polarity. With the emergence of new magnetic flux, the change of sheared angles of the horizontal magnetic field between the magnetic main poles of opposite polarity is insignificant in comparison with daily vector magnetograms, but the distribution of the intensity of the vertical current inferred from the horizontal magnetic field evolved only gradually. 3) The flare sites occurred near the magnetic islands and bays of opposite polarity and were associated with the change of the vector magnetic field. Although some of the flare sites are located near the peak areas of the vertical electrical current density, their corresponding relationship is insignificant.

\subsection{Magnetic Shear and Gradient}

The magnetic shear is an important parameter to measure the non-potentiality of magnetic field in solar active regions (cf. Severny, (1958), Hagyard et al. (1984), Chen et al. (1989), Chen et al. (1994), L\"{u}, Wang, \& Wang (1993), Schmieder et al. (1994), Wang et al. (1994), Zhang et al. (1994), Li et al. (2000), Wang et al. (2000)  
\cite{Sev58,Hagyard84,Chen89,Chen94,Lv93,Schmieder94,Wangh94,Zhang94,Li00,WangHN00}), 
while the non-potential field can also be measured from the strong magnetic gradient of active regions, which is strongly correlated with active region flare-CME productivity (cf. Severny (1958), Falconer (2001)  \cite{Sev58,Falconer01}. The change of vector magnetic fields during the solar flares was observationally presented, such as, by Chen et al. (1989) \cite{Chen89}.

\begin{figure*}[t]
\begin{center}
\includegraphics[width=170mm]{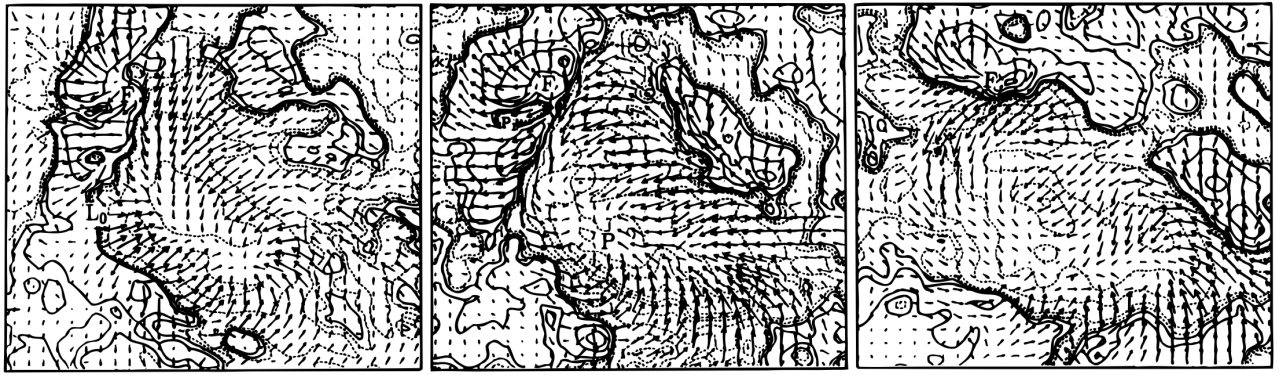}
\end{center}
\caption{Ambiguity-resolved vector magnetograms at 0050 UT on 1991 June 6, at 0353 UT on 1991 June 9, and at 0438 UT on 1991 June 12 (from left to  right) transformed from the image plane to heliographic coordinates. The contours indicate the longitudinal magnetic field distribution, $\pm$20, 160, 640, 1280, 1920, 2240, 2560, and 2880 G.    From \cite{Zhang96b} \label{fig:MGshear2}}
\end{figure*}

For analyzing the relationship between the non-potential magnetic field and electric current in solar active regions, the photospheric vector magnetograms, after the resolution of 180$^\circ$-ambiguity of transverse components of field,  in the $\delta$ active region NOAA 6659 in June  1991 are shown in Figure \ref{fig:MGshear2}, which observed at Huairou Solar Observing Station. A series of powerful flares occurred in this active region  as studied by Zhang (1996) \cite{Zhang96b}. It is found that the high shear of transverse magnetic field and gradient of  longitudinal magnetic field formed near the magnetic neutral line of the active region, where the transverse magnetic field rotated counterclockwise around the magnetic main pole of negative polarity. 

The shear angle can be weighted by the transverse magnetic field cf. Hagyard et al. (1984) \cite{Hagyard84}
\begin{equation}
\theta_T=B_{oh}\cdot \cos^{-1}\left(\frac{\textbf{  B}_{oh}\cdot\textbf{  B}_{ph}}{B_{oh}B_{ph}}\right),\label{eq:MGsh1}
\end{equation}
where $B_{oh}$ and $B_{ph}$ are the observed transverse field and that calculated from the magnetic charges in the approximation of potential field. The amplitude of the shear angle in eq. (\ref{eq:MGsh1}) reflects the non-potentiality of the active region.  

The horizontal gradient of the photospheric longitudinal magnetic field in active regions can be inferred from (cf. Leka \& Barnes (2003) \cite{Leka03}) 
\begin{equation}
|\bigtriangledown(B_z)|=\sqrt{\left(\frac{\partial B_z}{\partial x}\right)^2+
\left(\frac{\partial B_z}{\partial y}\right)^2}.\label{eq:MGsh2} 
\end{equation}
The main contribution of magnetic shear in the active regions comes from the deviation of the transverse field from the potential field inferred by magnetic charges in the photosphere, while the magnetic gradient comes from the non-uniformity of the longitudinal field.

As one knows  the non-potential magnetic field formed in solar active regions means the existence of electric current (cf. Moreton \& Severny, (1968) \cite{Moreton68}). The relationship between the magnetic field and electric current is   
\begin{equation}
\label{eq:current}
\textbf{  J}=\frac{1}{\mu_0}\bigtriangledown\times \textbf{  B},
\end{equation}
where $\textbf{  J}$ is in units of $\textrm{Am}^{-2}$ and $\mu_0=4\pi\times 10^{-3}\textrm{GmA}^{-1}$ is the permeability in free space.

\begin{figure*}[t] 
 \begin{center}
\includegraphics[width=50mm]{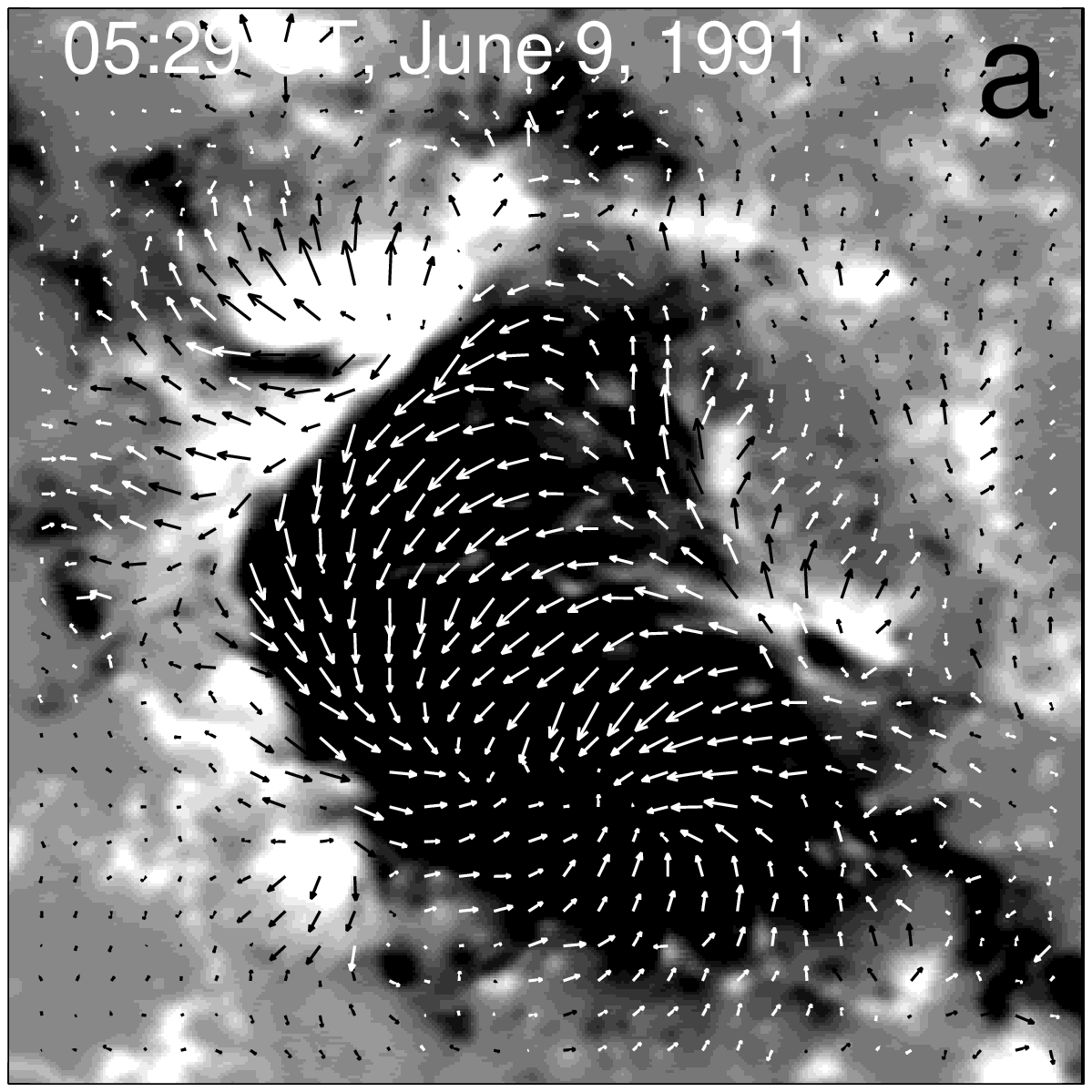} 
\includegraphics[width=50mm]{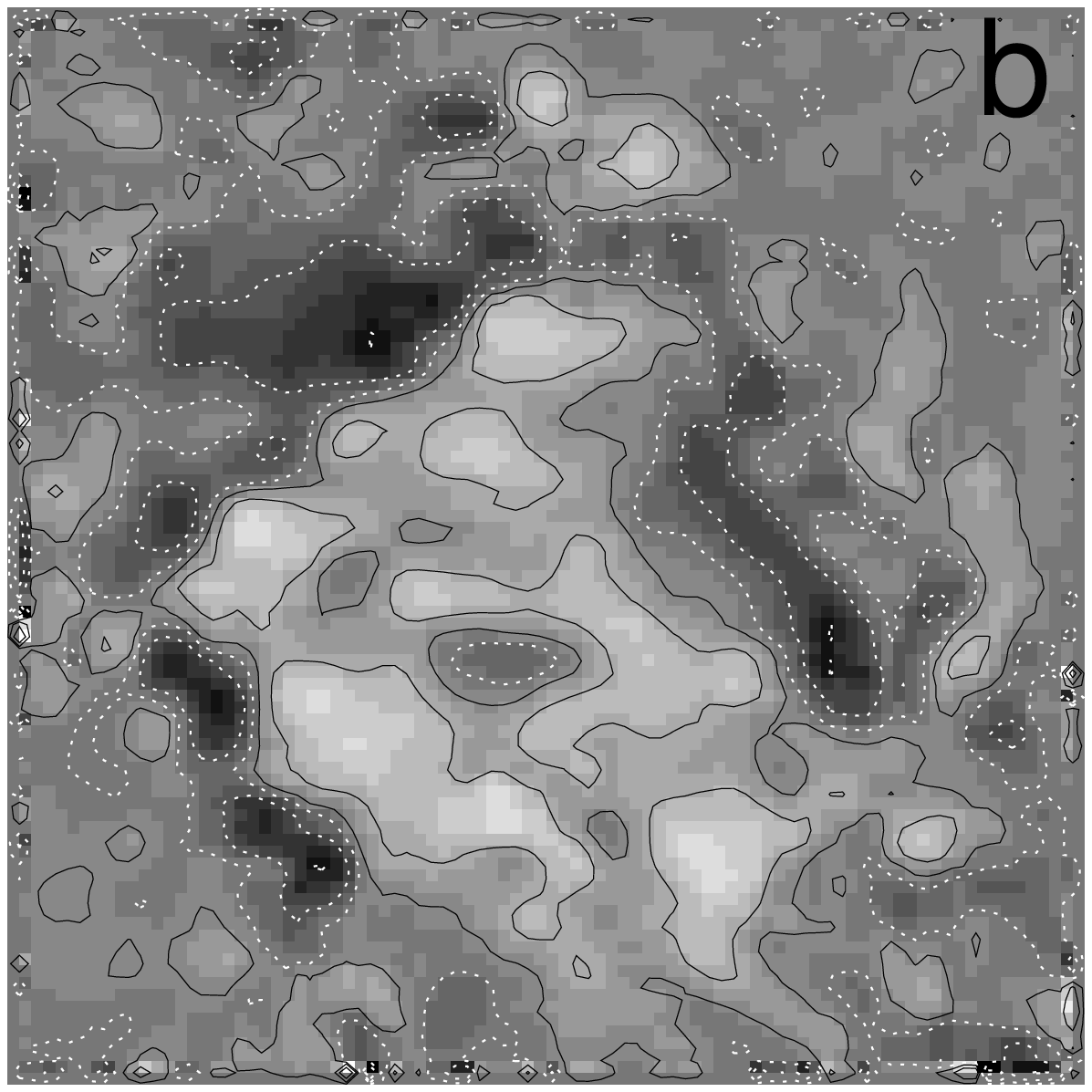} 
\includegraphics[width=50mm]{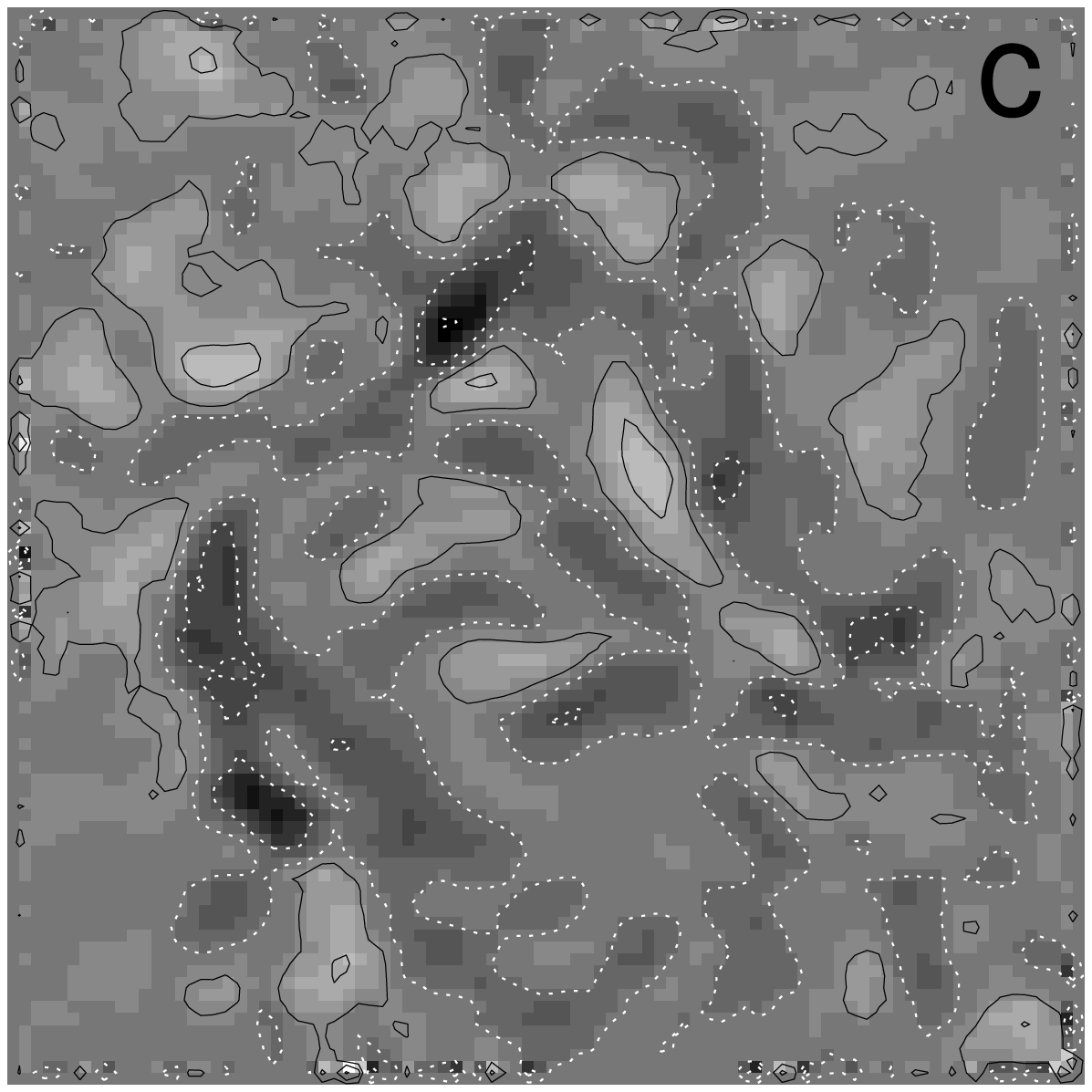}
\end{center}
\caption{(a) Photospheric vector magnetic field in the active region (NOAA 6659) at 05:29 UT on June 9, 1991. The white (black) shows the positive (negative) polarity and arrows mark the transverse field.    The corresponding vertical electric current (b) inferred by the term of the magnetic chirality of the magnetic field and (c)  by the term of the magnetic gradient and shear  term in Eq. (\ref{eq:towcurr}). The solid (dashed) contours correspond to the  upward (downward) flows of vertical currents of ${\pm}$ 0.002, 0.008, 0.02, 0.04, 0.075, 0.12 $\textrm{Am}^{-2}$. The size of view is $2.'72\times 2.'72$. \label{fig:sheargrad}}
\end{figure*}

As letting $\textbf{  B}=B\textbf{  b}$ and $\textbf{  b}$ is the unit vector along the
direction of magnetic field,
the current  may be written in the form \cite{Zhang01a}
\begin{equation}
\textbf{  J}=\frac{B}{\mu_0}\bigtriangledown\times \textbf{  b}+
\frac{1}{\mu_0}(\bigtriangledown B)\times\textbf{  b}.\label{eq:MGsh4}
\end{equation}
It is found that the electric current in solar active regions relates to the properties of chirality and gradient of magnetic field. The first term in eq. (\ref{eq:MGsh4}) connects with the twist of unit magnetic lines of force and intensity of the field. The second term in eq. (\ref{eq:MGsh4}) connects with the heterogeneity and orientation of  magnetic field.

The vertical electric current  can be written according to Eq. (\ref{eq:MGsh4})  
\begin{equation}
\label{eq:towcurr}
J_z=\frac{B}{\mu_0}\left(\frac{\partial { b}_y}{\partial x}-\frac{\partial {b}_x}{\partial y}\right)+\frac{1}{\mu_0}\left({ b}_y\frac{\partial B}{\partial x}-{b}_x\frac{\partial B}{\partial y}\right),
\end{equation}
where
${b}_x={{ B}_x}/{B}$ and ${b}_y={{B}_y}/{B}.$ The distribution of the two current components in active region NOAA6659  is shown in Figure \ref{fig:sheargrad}.  

Similar studies by means the two current components of active regions were taken, {\bf such as, on the resolution of the azimuthal ambiguity of vector magnetograms of solar active regions (Georgoulis et al. (2004) \cite{Georgoulis04}), and   the non-potentiality of solar active regions (Vemareddy (2017) \cite{Vemareddy17}). }

\subsection{Free Energy Contributed from Non-potential magnetic fields} 

Now we analyze the free magnetic energy contributed from different  components of magnetic fields (which is similar to Priest (2014) \cite{Priest14}, p117)  
\begin{eqnarray}
\label{ }
   W_o&= &\frac{1}{2\mu_0}\int B_o^2dV  = \frac{1}{2\mu_0}\int (B_n^2+2\textbf{  B}_n\cdot\textbf{  B}_p+B_p^2)dV\nonumber\\
   &=& \frac{1}{2\mu_0}\int (B_n^2+2\textbf{  B}_n\cdot\textbf{  B}_p)dV+W_p,
\end{eqnarray}
where $W_o$ is the observed magnetic energy, $W_p$ is the potential magnetic energy,  $B_o$ is the observed magnetic field, $B_n$ is the non-potential component of magnetic field and $B_p$ is the potential component of magnetic field, respectively.    

As we set $\textbf{  B}_p=\nabla\psi$, it is fund the free energy
\begin{eqnarray}
\label{ }
  W_f=  W_o-W_p
    =  \frac{1}{2\mu_0}\int B_n^2dV+\frac{1}{\mu_0}\oint_S \psi \textbf{  B}_n\cdot d\textbf{  s}.\nonumber      
\end{eqnarray}
If the vertical component $B_{ns}=0$ in the surface $S$ of the close volume, it is found 
\begin{equation}
   W_f= \frac{1}{2\mu_0}\int B_n^2dV.      
\end{equation}
While Zhang (2016) \cite{Zhang2016} noticed that the contribution of the parameter $\frac{1}{\mu_0}\textbf{  B}_{n}\cdot\textbf{  B}_{p}$  in the solar surface cannot be neglected normally in  the evolution of the flare-productive regions. This probably means the non-potential components of magnetic fields extend from the subatmosphere and bring the free energy into the atmosphere.


\begin{figure}[H] 
\centering
\includegraphics[angle=-90,scale=.32]{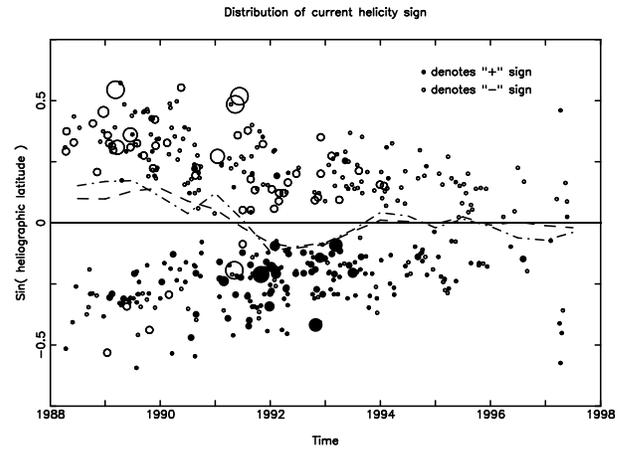}
\caption{Butterfly diagram of the electric current helicity. The mean density of the current
helicity of active regions is marked by size of the circles for grades: 0, 1, 3, 5, 7 ($10^{-3}\textrm{G}^2\textrm{m}
^{-1}$). The filled and hollow circles mark the negative and positive sign of helicity. The dash-dotted line marks the average value of current helicity and the dashed line marks the average value of the imbalance of current helicity after the data smooth. From \cite{ZhangBao98}}\label{fig:zb_1998}
 \end{figure}

\section{Distribution of Magnetic Helicity with Solar Active Cycles}

{ According to the formula of current helicity density (Bao \& Zhang, 1998) \cite{BZ98},
the vertical component of current helicity density in the photospheric lyaer can be inferred by photospheric vector magnetograms
\begin{equation}
h_{\bot c}=(\textbf{  B}\cdot\bigtriangledown \times \textbf{  B})_\bot\sim\alpha B^2,  
\end{equation}
where $\alpha=\textbf{  b} \cdot\bigtriangledown \times \textbf{  b}$.\label{eq:MGsh7}}
   
In addition to the solar flare activities with helicity, such as, Bao et al. (1999), Zhang et al. (2008), Yang et al. (2012) \cite{Bao99,ZhangY08,YangX12},  it is found that the mean magnetic helicity density in the solar active regions tends to show negative signs in the northern hemisphere  and positive ones in the southern hemisphere (Seehafer (1990), Pevtsov, Canfield, \& Metcalf (1995), Abramenko et al. (1996), Bao \& Zhang(1998), Zhang \& Bao (1998), Hagino \& Sakurai (2004), Hao \& Zhang (2011) \cite{Seehafer90,Pevtsov95,Abramenko96, BZ98, ZhangBao98,Hagino04,Hao11} ). Figure \ref{fig:zb_1998} shows the distribution of mean current helicity density of active regions in 1988 - 1997 inferred from vector magnetograms observed at Huairou Solar Observing Station. Moreover, the reversal sign of helicity in these statistical work has also attracted much attention \cite{zetal10b}. 
The change of magnetic helicity with solar cycles, such as helicity butterfly diagram, also bring some new thoughts on the generations of magnetic fields with the dynamo process in the subatmosphere (such as, Kleeorin et al. (2003), Zhang et al. (2012), Pipin et al. (2013) \cite{Kleeorin03,zetal12,Pipin13}).

\section{The question on accuracy of measured vector magnetic fields}

\begin{figure}[H] 
\includegraphics[angle=0,scale=0.57]{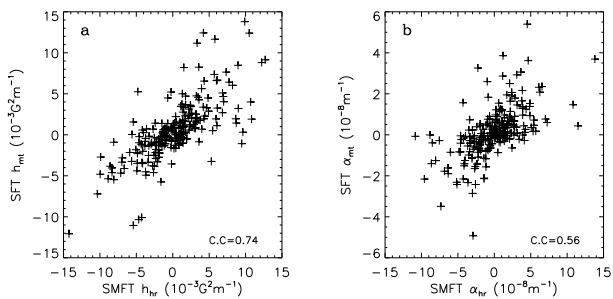}
\caption{Correlations in (a) $h_{\rm c}$ and (b) $\alpha_{\rm av}$ obtained
from SMFT (Huairou) and SFT (Mitaka) data. The values of correlation coefficients (CC)
are shown on the panels. From \cite{Xu16}}\label{fig:xu2}
 \end{figure}

In addition to some fundamental achievements on discovering some important properties of solar magnetic activities, it is also needed to notice that the magnetic fields and relevant   parameters, such as helicity etc.,  observed with different instruments not only have shown the basic same tendency, but also with some differences \cite{WangHM92,Bao00, Pevtsov06, Xu16}. As a sample, the  discrepancies on the statistical distribution of the current helicity parameters $h_c$ and $\alpha$ of solar active regions obtained by different solar vector magnetographs at Huairou in China and Mitaka in Japan can be found by Xu et al. (2016) \cite{Xu16} in Figure \ref{fig:xu2}.

This relates to a basic question on the accuracy for the measurements of the solar magnetic fields by means of solar vector magnetographs and the corresponding observing theories of solar magnetic fields. Some restrictions on the calibration of magnetic field from Stokes parameters have been presented due the nonlinearity and simplicity  in the above  calculation of the radiative transfer equations of the spectral lines in solar magnetic atmosphere. 
 It means that there are still some questions on the theories of the solar observations relative to the diagnostic with the radiative transfer of spectral lines, even if some problems of solar instrumental techniques have been ignored.  
 
\section{Discussions for some new challenges on measurements of solar magnetic fields}

We have introduced the fundamental background of the radiative transfer process of the magnetic sensitive lines in the solar magnetic atmosphere, and corresponding analysis of solar vector magnetograms based on the study at Huairou Solar Observing Station of National Astronomical Observatories, Chinese Academy of Sciences.

As a series of vector magnetograms have been observed and some relevant interesting results have been presented from these observations, we should notice that the meticulous fine analysis based on the relative complex calculated from observed magnetograms probably {\bf is still needed,} even if some results probably are more interesting, such as the meticulous configuration and variation of the magnetic shear, free energy density, electric current and helicity in the solar active regions.  We probably also need to keep in mind that the quantitative analysis of solar magnetic field from observed magnetograms contains some intrinsic difficulties. 

This means that there are still some basic questions on the measurements of solar magnetic fields need to be carefully analyzed. Some of them have been presented in the following:

\begin{itemize}

   \item Radiative transfer of magnetic sensitive lines: 

We have presented the radiative transfer equations of magneticlly sensitive lines under the different approximations， whether ignoring the scattering or the assumption of thermodynamic  equilibrium, such as the numerical and analytical forms  (in subsec. \ref{se:rtsp}), which bring some ambiguity for the accurate  results.

The analytical solution  has normally been used to inverse the photospheric vector magnetic fields. This means that the varieties of solar atmosphere models, such as the sunspot, facula and quiet Sun, have been ignored. From the calculated results, we can find the different magnetic sensitivities of polarized spectral lines for the different kinds of solar atmospheres, such as the quiet Sun and sunspots in Figure \ref{fig:moline5324fra}.  It is also noticed that the different sensitivity for the measurements of the longitudinal and transverse components of magnetic field can be easily found from Eqs. \eqref{eq:weakapproxa}, i.e. the longitudinal and transverse fields relate to the first and second order derivative of the profiles of spectral lines, respectively.       
In the condition Eqs. \eqref{eq:weekfild1} and (\ref{eq:weekfild2}) of the approximation of the weak fields, the influence of magneto-optical effects has been neglected for the transverse components of the magnetic fields obviously. We can find that these   provide the constraint condition on the observations of solar magnetic fields by different approximate methods or situations.

  \item Solar model atmosphere:  

We have presented the observations of photospheric vector magnetic fields with the FeI$\lambda$5324.19\AA{} line and chromospheric magnetic fields with the H$\beta$$\lambda$4861.34\AA{} line based on the analysis of the radiative transfer equations for polarized radiation  by Unno (1956) \cite{Unno56} and  Rachkovsky (1962a, b)  \cite{Rachkovsky62a,Rachkovsky62b} under the assumption of  local thermodynamic equilibrium or their extension to non-local thermodynamic equilibrium at Huairou Solar Observing Station. 

Even if the numerical calculation can be used in the analysis of the radiative transfer equations (\ref{radtrf})  of Stokes parameters and provides some important information on the formation of polarized spectral lines in the different atmosphere models, it still rarely has been used in the real inversion of the magnetic field, due to some difficulty and arbitrariness in the accurate selections of inputed atomic and solar parameters.  

It is noticed that the different formation heights of magnetic sensitive lines occur in the quiet Sun and sunspots, such as the Wilson effect of sunspots,  due to the the transparency of sunspots, for the measurements of photospheric magnetic field,  and the different heights for different chromospheric features, such as the prominence, fibrils and plages etc.  This means that the observed magentograms probably does not always provide the  information of magnetic fields at the same height in the solar atmosphere, even if at the similar optical depths of the working spectral lines.

{ A notable question is the detection of the configuration of magnetic fields in the solar eruptive process, cf. Chen et al. (1989) \cite{Chen89}. Some  study  with Stokes parameters of spectral lines can be found  (Hong et al.(2018) \cite{Hong18}).  Due to the distortion of the spectral lines with the variation of solar atmosphere in the solar eruptive process \cite{Fang08}, the analysis of relevant Stokes parameters of the spectral lines is a notable topic. }

  \item Measurements of magnetic fields in the corona: 
  
The Hanle effect is a reduction in the polarization of light when the atoms emitting the light are subject to a magnetic field in a particular direction, and when they have themselves been excited by polarized light. The Hanle effects with polarized light in the corona is notable question for the measurements of the coronal magnetic fields, cf. Liu \& Lin (2008), Qu et al. (2009), Li et al. (2017) \cite{LiuLin08,Qu09, Li17}. 

The analysis of the polarized lights due to the Hanle effect for the diagnostic of the coronal magnetic field is still a challenging topic \cite{Stenflo94}. 
  
  \item 180$^\circ$ ambiguity on the determination of  transverse magnetic fields: 

The resolution of the 180$^\circ$ ambiguity of transverse fields is a difficult question due to the property of polarized light with Zeeman effects.  

Several  basic assumptions and approaches have been used to resolve the 180° ambiguity of transverse components of magnetic fields, such as comparing the observed field to a reference field or direction, minimizing the vertical gradient of the magnetic pressure, minimizing the vertical current density, minimizing some approximation to the total current density, and minimizing some approximation to the field's divergence (Metcalf et al. (2006) \cite{Metcalf06}, Georgoulis (2012) \cite{Georgoulis12}). Which of these treatments is prioritized is still debatable or questionable.

The projection of vector magnetic fields from solar disk to the heliospheric coordinates relates the transform of the different components of magnetic field \cite{Hagyard87}, such as in Figure \ref{fig:MGshear2}, while the  different Stokes parameters relate to different sensitivities and noise levels (cf. Eqs. \eqref{eq:weakapproxa}). This leads to the inconsistency in the  transformation of different components of magnetic fields. It also causes degradation in spatial resolution of the vector magnetic fields after  the inevitable smoothing fields from the image plane to heliographic coordinates in Figure \ref{fig:MGshear2}.

  \item Diagnostics of electric fields in solar atmosphere: 
  
The  Stark effect is notable in the broadened wings of  hydrogen  and helium lines. It has been used in the analysis of the radiative transfer of H$\beta$ line for the measurements of magnetic fields at Huairou Solar Observing Station, see Eqs. \eqref{eq:HB2} and \eqref{eq:HB3}.
  
The Stark effect is the shifting and splitting of spectral lines of atoms and molecules due to the presence of an external electric field. 
Normally, the isotropic plasma is statistically electrically neutral, due to the  Debye effect. The electric fields in the solar flares were presented by Zhang \& Smartt (1986)  \cite{ZhangSmartt86} based on the spectral analysis of the linear and quadratic components of the Stark broadening. 

A dynamic and quantitative depiction of the induction electric field $\textbf E=\textbf u\times \textbf B$ on the changes of the active regions  was described by Liu et al. (2008) \cite{Liu08}, where $\textbf u$ is the velocity of the footpoint motion of the magnetic field lines and $\textbf B$ is the magnetic field. $\textbf E$ represents the dynamic evolution of the velocity field and the magnetic field, shows the sweeping motions of magnetic footpoints, exhibits the buildup process of current, and relates to the changes in nonpotentiality of the active region in the photosphere.
 
The measurements of the polarized light due to the Stark effect (or electric fields) in the anisotropic plasma is also a challenging topic.

 \item   Horizontal component of electric current and corresponding helicity:

The observations of chromospheric magnetic field { are important for diagnosing the spatial configuration of magnetic lines of force as compared} with the photospheric one, while the disturbance of photospheric blended lines in the wing of H$\beta$ line also bothers the detection  of the magnetic field in the high solar atmosphere in Figure \ref{fig:StokesHV1}. The lower sensitivity of Stokes $Q$ and $U$ of the H$\beta$ line cripples the observations of the transverse components of chromospheric magnetic fields.

The electric current helicity density $h_c=\langle \varepsilon_{ijk}b_i{\partial b_k/\partial x_j}\rangle$ contains six terms, where $b_i$ are components of the magnetic field. Due to the observational limitations, only four of the above six terms can be inferred from solar photospheric vector magnetograms. { By comparing the results for simulation, Xu et al. (2015) \cite{Xu15} distinguished the statistical difference of above six terms for isotropic and anisotropic cases.  
This means that the electric current and magnetic (current) helicity inferred from the photospheric vector magnetograms do not contain the completeness in theoretical sense. } 

  \item The limited spatial resolution for the measurements of magnetic fields: 

  It is normally believed that the mean free path of photons emerging from the solar photosphere and chromosphere is about 100 km (Zuccarello (2012), Judge et al. (2015)  \cite{Zuccarello12,Judge15}). This relates to the detection of the magnetic flux tules in the solar atmosphere and their possible distribution.

  \item Influence of the Doppler motion for the measurements of solar magnetic  fields: 

The influence for the measurements of  magnetic fields from the Doppler velocity field due to the solar rotation was discussed (such as  by  Wang, Ai, \& Deng (1996) \cite{WangTJ96}), and some for Doppler measurements of the velocity field in the solar photosphere and implications for helioseismology can be found  (Rajaguru et al. (2006) \cite{Rajaguru06}).  However, the influence on the accuracy of the magnetic field measurement for the moving objects in the solar atmosphere is still a subject worth exploring. 
  
\end{itemize}

Based on the above discussions, we can find that the diagnostic of solar magnetic fields in the solar atmosphere still is a fundamental topic.
{ Although we have made some achievements on the analysis of solar vector magnetic fields based on the theory of the radiative transfer of spectral lines from observations, the quantitative study and the corresponding assessment on its accuracy still remain some to be made, especially on the aspects of exact inference from vector magnetograms, such as the electric current and helicity etc, due to some theoretical ambiguities and errors, even if one excludes the questions caused from the observational techniques. }
Comparison on the vector magnetograms observed by different magnetographs probably is one of the advisable methods  (Wang et al. (1992), Bao et al. (2000), Zhang et al. (2003), Liang et al. (2006), Xu et al .(2016) \cite{WangHM92, Bao00, zetal03, Liang06, Xu16}).

In this paper, we have taken more attention on the measurements of magnetic fields at Huairou Solar Observing Station. Of course, there are still more questions on the measurements of solar magnetic fields that we have not covered yet or discussed in detail.

\section{Acknowledgements}
This study is supported by grants from the National Natural Science
Foundation (NNSF) of China under the project grants 11673033, 11427803, 11427901 and Huairou Solar Observing Station, Chinese Academy of Sciences. 

\end{multicols}
\end{document}